\documentclass{article}
\usepackage[utf8]{inputenc}
\usepackage{hyperref}
\usepackage{arxiv}

\usepackage{graphicx,amsmath,amsfonts,amscd,amsthm,amssymb,pstricks}
\usepackage{color}
\usepackage{dcolumn}
\usepackage{bm}
\usepackage{longtable}
\usepackage{mathrsfs}
\usepackage{textcomp}
\usepackage{esvect}
\usepackage{float}
\usepackage[british]{babel}
\usepackage{environ}
\usepackage{xifthen}
\usepackage{xargs}		
\usepackage{physics,mathtools}
\usepackage[separate-uncertainty = true]{siunitx}
\usepackage{multirow}
\usepackage{comment}
\usepackage{setspace}
\usepackage{apacite}
\newcommand{\ham}{\mathcal{H}}

\def\block(#1,#2)#3{\multicolumn{#2}{c}{\multirow{#1}{*}{$ #3 $}}}
\allowdisplaybreaks[1]

\newtheorem*{theorem*}{Theorem}
\newtheorem*{lemma*}{Lemma}

\usepackage[sort&compress]{natbib}

\usepackage{graphicx, color}
\graphicspath{{fig/}}

\usepackage{algorithm, algpseudocode} 
\usepackage{mathrsfs} 

\title{Single-particle Non-linear Beam Dynamics}
\author{M. Giovannozzi\\
Beams Department, CERN, Esplanade des Particules 1, 1211 Meyrin, Switzerland}
\begin{document}
	\maketitle
	
\begin{abstract}
The investigation of single-particle dynamics in circular particle accelerators can be traced back to pioneering works in the 1950s. Traditionally, the design of new circular accelerators has focused on optimising linear dynamics, striving to minimise non-linear effects. When encountered, these effects were managed using suitable correction techniques. However, in recent years, this methodology has undergone significant reconsideration, leading to a more favourable view of non-linear beam dynamics. Notably, several new concepts have been promoted to improve the description and modelling of non-linear beam dynamics. Furthermore, new proposals have emerged that leverage the extensive potential of non-linear beam dynamics to enhance the control and manipulation of the characteristic features of charged particle beams. This article provides a detailed review and discussion of all these innovative approaches.

	\noindent\textbf{Keywords:} Circular colliders and storage rings; dynamic aperture; diffusive models; stable islands; adiabatic trapping and transport; transition-energy jump
\end{abstract}

	
\section{Introduction}
Although studies on devices to accumulate and accelerate beams of charged particles date back to the beginning of the XX\textsuperscript{th} century, it was with the advent of modern circular accelerators, colliders, and storage ring in the 1950s that the seminal papers of~\cite{Christophilos,PhysRev.88.1197,PhysRev.88.1190,Courant:593259} were written. These derived the strong-focussing, or alternating-gradient, principle and set the foundations for the description of the motion of a charged particle in a magnetic lattice made of dipoles and quadrupoles. According to this theory, the motion of charged particles in the horizontal or vertical planes, also called betatron motion, is represented by harmonic oscillations whose frequency, called tune, is a property of the magnetic lattice and whose amplitude is modulated along the circumference of the accelerator. At its foundation, the beam dynamics is linear in the sense that the superposition principle applies to the solutions of the differential equations describing the motion of the charged particles.

The unique departure from the linear paradigm is represented by the use of sextupole magnets in accelerator lattices. These devices, generating a field quadratic in the transverse coordinates, are needed to provide control of the linear chromaticity, i.e. the dependence of the tune on the momentum of each charged particle. A second application of sextupole magnets is to extract particles over millions of turns using so-called slow extraction (see, e.g. Refs.~\cite{Tuck:1951,LeCouteur:1951,gordon:1958,hammer:1961,kobayashi:1967,gordon:1971} for an account on the first stages in the domain). This was the first application of non-linear resonances to circular accelerators. The effect of the sextupole magnets is normally taken into account by assuming that they generate a perturbation of the linear dynamics, which remains the reference concept. 

A considerable paradigm shift occurred at the time of the design studies for the modern generation of colliders (see, e.g.~\cite{RevModPhys.93.015006} for a review) based on superconducting magnets (see, e.g.~\cite{Tollestrup:1141042} and references therein), such as the Tevatron~\cite{edwards:1985,shiltsev:2012,holmes:2013,denisov:2022} at Fermilab, the Hadron Elektron Ring Anlage (HERA)~\cite{voss:1994} at DESY, the Relativistic Heavy Ion Collider (RHIC)~\cite{hahn:2003,harrison:2003} at BNL, and the Large Hadron Collider (LHC)~\cite{LHCDR,evans:2012,myers:2013} at CERN. The inherent principle on which superconducting magnets are based, i.e. that the field properties are generated by the distribution of currents\footnote{In the case of classical magnets the shape of the iron pole controls the properties of the magnetic field.}, implies that non-linear magnetic field errors are unavoidable. As a reaction to this new situation, the efforts were directed along two main research lines: The description of the non-linear beam dynamics with the goal of determining efficient indicators or methods to describe the beam evolution and hence infer the actual accelerator performance in the presence of the non-linear magnetic field errors; the study of efficient techniques to compensate for the non-linear errors by means of a set of corrector magnets judiciously located in the accelerator lattice. Of course, these approaches are highly interconnected and complementary.

The research line on the improved description of the non-linear beam dynamics focused on the characterisation of the dynamics, whose orbits are distorted with respect to the linear case with resonances excited, resulting in chaotic motion and unbounded orbits. The first consequence led to the consideration of Lyapunov exponents~\cite{benettin:1980} as suitable tools to inspect the regular or chaotic character of orbits. This technique was implemented in tools for numerical simulation of beam dynamics to study its reliability (see, e.g. Refs.~\cite{Schmidt:1991ha,dynap1}). The second consequence was studied by inspecting the boundedness of orbits by means of numerical simulations to compute the so-called dynamic aperture (DA), which represents the extent of the phase-space region where the orbits remain bounded over a finite time. By carefully considering phase-space sampling~\cite{PhysRevE.53.4067} and applying the appropriate averaging procedure, it was possible to determine the time evolution of the DA~\cite{PhysRevE.57.3432,Bazzani:2019csk} and to describe it using a robust model based on the stability-time estimate from the Nekhoroshev theorem~\cite{Nekhoroshev:1977aa,Bazzani:1990aa,Turchetti:1990aa}. The methodology for DA analysis that uses the scaling law regarding the time evolution of DA has recently been improved through the integration of machine learning techniques. This aims to optimise the quality of the fit of the analytical model with numerical data~\cite{MDPI_ML,casanova:2023}. It is worth stressing that the approach based on DA scaling laws opened new opportunities, such as the prediction of the DA over time scales that are compatible with the storage time of colliders, in spite of these time scales being well beyond the capabilities of direct numerical simulations. Furthermore, it allowed the establishment of a direct link between DA and intensity evolution in circular accelerators~\cite{PhysRevSTAB.15.024001}, improved models of luminosity evolution in hadron colliders~\cite{Giovannozzi:2018wmm,Giovannozzi:2018igq}, and an innovative method for measuring DA~\cite{PhysRevAccelBeams.22.034002}. 

A complementary approach considered the development of diffusion models to describe the evolution of the transverse beam distribution (see, e.g. Refs.~\cite{cary:pac87,Brüning:1108270,Zimmermann:1994ec,bruening:epac94,zimmermann:1995,Turchetti:1120253,Bazzani:1120261} and references therein), an activity that also included extensive experimental programmes at various accelerators around the world~\cite{fischer:1995,PhysRevE.55.3507,PhysRevSTAB.16.021003,PhysRevAccelBeams.23.044802,PhysRevLett.68.33,gerasimov1992applicability,MESS1994279,Zimmermann:1994ec,PhysRevLett.77.1051,PhysRevSTAB.5.074001,stancari2011diffusion,PhysRevSTAB.15.101001}. The underlying assumption is that non-linear magnetic field errors, combined with all sorts of time-dependent effects, such as ripple in the power converters of the magnetic devices, induce a large-scale chaotic region. In this region, the dynamics closely resembles a diffusive process, and the features of this process are folded in the functional form of the diffusion coefficient. In most of the studies, phenomenological assumptions were made and in the experimental studies the functional form was not assumed \textsl{a priori}, but a local approach was used. This consisted of carrying out measurements of the diffusion coefficients at various amplitudes trying to derive \textsl{a posteriori} a functional form of the diffusion coefficient as a function of the amplitude in phase space. A different approach has been proposed recently, in which, once again, the Nekhoroshev theorem has been invoked to support a specific functional form of the diffusion coefficient. This approach has also been tested in experiments with extremely encouraging results~\cite{Bazzani:2019lse,bazzani2020diffusion,montanari:ipac2021:tupab233,our_paper9,montanari:ipac22-mopost043,montanari:ipac2023-wepa022,montanari:2025,Montanari:2863942}.   

Then there was the research activity aimed at developing effective correction methods to reduce the impact of magnetic field errors on the dynamics of single-particle beams (refer to Refs.~\cite{neuffer:1987,neuffer:1989,neuffer:1989a,scandale:1990} for examples). The overarching goal was to identify an observable closely linked to the non-linear nature of beam dynamics, which could serve as a quantitative criterion for determining the necessary strength of correctors for the magnetic lattice. One can see how this pursuit aligns well with the first research direction. Local and global correction strategies were evaluated. A sample of all these efforts is documented in Refs.~\cite{wei:pac99-tha143,pilat:epac02-weple042,pilat:pac03-tppb041,pilat:pac05-woac007,luo:pac07-frpms110,zimmer:pac11-thp065}. Moreover, it is notable that there was a keen interest in integrating these findings with a rigorous experimental programme in currently operating circular colliders (see, e.g. Refs.~\cite{PhysRevSTAB.18.121002,PhysRevAccelBeams.22.061004,maclean:2022,PhysRevAccelBeams.26.121001}). To accurately portray the context of these activities, it is essential to recognise that much of this work was inspired by the US's substantial efforts for the ultimately unsuccessful Superconducting Super Collider (SSC) project (see Refs.~\cite{wojcicki:2008,wojcicki:2009} and references therein) and that during that period, there was a renewed and significant interest in adapting and evolving concepts and methodologies from the theory of dynamical systems to the field of non-linear beam dynamics (refer to Refs.~\cite{dragt:1976,dragt:1979,cary:1981,dragt:1983,dragt:1987,forest:1987,bazzani:1988,forest:1989,scandale:1991,bazzani:1993,Dragt:2011vea} for further details). These endeavours have proven to be highly rewarding for the latest advancements, where non-linear phenomena are now examined with a fresh and radically different perspective.

At the dawn of the XXI\textsuperscript{th} century, the need to replace the traditional lossy type of extraction for high-intensity proton beams~\cite{bovet73} from the Proton Synchrotron (PS) to the Super Proton Synchrotron (SPS) triggered a new line of research that made it possible to fully exploit the immense possibilities offered by non-linear dynamics. The new technique was based on the generation of stable resonances through the use of sextupolar and octupolar magnets combined with an appropriate time variation of certain accelerator parameters (in general the transverse tune of the accelerator)~\cite{PhysRevLett.88.104801}. If the variation is slow, so that it can be considered adiabatic, then when the stable islands move in phase space they  capture the charged particles they encounter in their movement. This enables splitting a single beam in the transverse plane so that the initial single-Gaussian beam distribution is replaced by a final multi-Gaussian distribution, with the number of Gaussian distributions at the end of the splitting process a function of the order of the resonance.  In addition to trapping in the islands, it is possible to transport the trapped particles to large amplitudes. All this can be done without any kind of beam loss. This type of beam manipulation combines the use of resonances with the theory of adiabatic trapping~\cite{neish1975,NEISHTADT198158,NEISHTADT1986,an4,NEISHTADT1991,an6,an9,an10,Neishtadt2013,Neishtadt_2019}, which is a powerful tool to create new types of beam manipulation schemes. The first example was this Multi-turn extraction (MTE) developed at the PS which, after a long period of study and experimental testing~\cite{PhysRevSTAB.7.024001,Giovannozzi:987493,PhysRevSTAB.9.104001,PhysRevSTAB.12.014001}, reached maturity and operational use in the second half of 2015~\cite{Borburgh:2137954,PhysRevAccelBeams.20.061001,PhysRevAccelBeams.20.014001,PhysRevAccelBeams.22.104002}. This technique has continued to be developed adding improvements and now also combines sophisticated manipulation of the longitudinal beam distribution, achieved with pulsed radio frequency cavities that create a barrier bucket~\cite{Vadai:IPAC19-MOPTS106,Vadai:IPAC19-MOPTS107,Vadai:2702852,PhysRevAccelBeams.25.050101}.  

MTE is the prototype of a series of new manipulations based on the use of non-linear dynamics and adiabatic theory. It is possible to devise a new injection technique, Multi-turn Injection (MTI)~\cite{PhysRevSTAB.10.034001}, by applying a time inversion to MTE, which provides an elegant approach to shape the transverse beam distribution. Furthermore, beams with multi-mode transverse distributions could be used to mitigate electron-cloud effects~\cite{Cui:2708898}. If, instead of using a one-dimensional resonance, a two-dimensional resonance is used, it is possible to manipulate transverse emittances, violating the conservation of emittances that applies in the linear case, in a completely controlled manner~\cite{PhysRevAccelBeams.24.094002,our_paper7,PhysRevAccelBeams.25.104001,Capoani:2863096}. One of the most recent developments is the consideration of oscillating magnetic-field devices (dipolar or non-linear field type), with which it is possible to manipulate the transverse distribution of the beam and to split it into several Gaussian functions~\cite{our_paper4}. It has recently been possible to use this approach to cool an annular distribution~\cite{PhysRevAccelBeams.26.024001} and show that these elements can be used to clean the halo of a beam~\cite{capoani:2025}. 

The latest development is the combined use of stable islands and bent crystals to achieve a new technique to generate a slow beam extraction from a circular accelerator~\cite{PhysRevResearch.6.L042018}. 

In addition to MTE 
almost all other manipulations relying on non-linear dynamics that have been mentioned above are still at the stage of theoretical studies complemented by numerical simulations on Hamiltonian models or simple models of accelerator lattices. The exception to this is the manipulation of transverse emittances using the crossing of a two-dimensional resonance, which is currently being studied experimentally at the PS ring. 

As a final remark, it should be noted that the presence of stable islands in phase space generates additional closed orbits to the standard orbit at the origin of the phase space. The optical properties around each closed orbit are different, which opens up the possibility of multiple beams experiencing different optics. This can be used to create extraction schemes without septum magnets~\cite{PhysRevSTAB.18.074001} for example, or sophisticated schemes to cross the transition energy in a non-adiabatic way~\cite{GiovannozziEPJPlusTransition}. The latter development is being studied with the intention of applying it to the Electron Ion Collider (EIC)~\cite{eic-cdr} for which numerical studies and experiments are underway at RHIC~\cite{lovelaceiii:ipac23-mopa056,lovelaceiii:ipac24-tups03}. 

Following the domain overview, it is important to underscore that the primary aim of this article is to examine the single-particle non-linear beam dynamics in circular accelerators, where the particle's behaviour is characterised by a time-independent Hamiltonian function, in which energy damping or stochastic processes due to photon emission can be neglected. This context is most applicable to hadron accelerators. Conversely, in lepton circular accelerators, the beam dynamics is significantly influenced by energy damping and quantum emission processes, even at moderate values of the beam energy. In this situation, the particle's energy is not constant and its motion is inherently stochastic. Despite these differences, non-linear beam dynamics also offers innovative possibilities to enhance beam manipulation techniques for lepton machines, potentially improving the performance of circular lepton colliders. Though not comprehensive, most of the existing research concerning non-linear beam dynamics in lepton accelerators is elaborated in Refs.~\cite{Ries:IPAC2015-MOPWA021,goslawski:ipac16-thpmr017,goslawski:ipac17-wepik057,kramer:ipac18-tupml052,Goslawski:IPAC2019-THYYPLM2,tavares:ipac19-tuyplm3,holldack:2020,olsson:2021,arl:ipac22-thpopt003,kim:ipac22-mopost053,wang:ipac22-mopost051,franchi:2022,wang:ipac23-wepl093,PhysRevAccelBeams.26.104001,wang:ipac24-mops05}. The extension of the theoretical framework developed for the Hamiltonian case, which has been extensively studied both theoretically and experimentally in hadron accelerators, is now being considered to provide a solid basis for novel beam manipulation in lepton accelerators.  

The outline of this article is the following: In Section~\ref{sec:desc}, key concepts to describe and model the single-particle beam dynamics are discussed, namely the dynamic aperture and the use of diffusive processes to describe the non-linear beam dynamics. Section~\ref{sec:expl} introduces and thoroughly examines the innovative framework designed to leverage the extensive opportunities provided by the complex characteristics of non-linear beam dynamics. Lastly, conclusions are drawn in Section~\ref{sec:conc}.
\section{Describing and modelling non-linear beam dynamics} \label{sec:desc}
\subsection{Dynamic aperture} \label{sec:da}
The concept of dynamic aperture (DA) is essential to describe the impact of the non-linear effects on the beam dynamics. The DA represents the extent of the phase space volume in which the orbits remain bounded over a finite lapse of time. Assuming boundedness for a finite duration is crucial to prevent mathematical anomalies, because extending to infinite time results in the DA reducing to zero, influenced by topological effects like Arnold diffusion~\cite{arnold:1964}.

Following~\cite{PhysRevE.53.4067}, let us consider the phase-space volume of the initial conditions that are bounded after $N_\mathrm{t}$ turns around the ring circumference, namely
\begin{equation}
\int \int \int \int \chi(x_1,p_{x_1},x_2,p_{x_2}) \; \dd x_1 \, \dd p_{x_1}  \, \dd x_2 \, \dd p_{x_2} \, ,
\label{eqn8o}
\end{equation}
where $\chi(x_1,p_{x_1},x_2,p_{x_2})$ is the generalisation of the characteristic function to the 4D case, i.e. it is equal to one if the orbit starting at $(x_1,p_{x_1},x_2,p_{x_2})$ is bounded, or zero otherwise. 

To exclude the disconnected part of the stability domain in the integral~\eqref{eqn8o}, a suitable coordinate transformation should be chosen. Since linear motion is the direct product of constant rotations, the natural choice is to use the polar variables $(r_i,\vartheta_i)$, where $r_1$ and $r_2$ are linear invariants. The non-linear part of the equations of motion adds a coupling between the two planes, the perturbative parameter being the distance to the origin. Therefore, it is natural to replace $r_1$ and $r_2$ with the polar variables $r \cos \alpha$ and $r \sin \alpha$, respectively, and the final form of the coordinate transformation reads
\begin{equation}
\left \{ \begin{array}{lcll}
  x_1 &=& r \cos \alpha  \cos \vartheta_1  &           \\
p_{x_1} &=& r \cos \alpha  \sin \vartheta_1  &      \qquad \qquad r \in [0,+\infty[             \\
  & &  &           \qquad \qquad \alpha \in [0,\pi/2]      \\
  x_2 &=& r \sin \alpha  \cos \vartheta_2  &           \qquad \qquad \vartheta_i \in [0,2\pi[ \qquad i=1,2      \\
p_{x_2} &=& r \sin \alpha  \sin \vartheta_2    \, ,  &        
\end{array} \right .
\label{eq00o}
\end{equation}
and substituting in Eq.~\eqref{eqn8o} we obtain
\begin{equation}
\int_0^{2\pi} \int_0^{2\pi} \int_0^{\pi/2}\int_0^\infty 
\; \chi(r, \alpha, \vartheta_1, \vartheta_2) \, r^3 \sin \alpha \cos \alpha \; \dd \Omega_4 \, , 
\end{equation}
where $d\Omega_4$ represents the volume element
\begin{equation}
\dd \Omega_4 = \dd r \, \dd \alpha \, \dd \vartheta_1 \, \dd \vartheta_2 \, .
\end{equation}

Having fixed $\alpha$ and $\boldsymbol{\vartheta}=(\vartheta_1,\vartheta_2)$, let $r(\alpha, \boldsymbol{\vartheta},N_\mathrm{t})$ be the last value of $r$ whose orbit is bounded after $N$ iterations. Then, the volume of a connected stability domain is expressed by 
\begin{equation}
A_{\alpha,\boldsymbol{\vartheta},N_\mathrm{t}} = \frac{1}{8} \, \int_0^{2\pi} \int_0^{2\pi} \int_0^{\pi/2} [r(\alpha,\boldsymbol{\vartheta},N_\mathrm{t})]^4 \sin 2 \alpha \; \dd \Omega_3 \, ,
\label{eqn3o}
\end{equation}
where
\begin{equation}
\dd \Omega_3 = \dd \alpha \, \dd \vartheta_1 \, \dd \vartheta_2 \, ,
\end{equation}
which allows excluding stable islands that are not connected to the main stable domain. We then define the DA as the radius of the hypersphere that has the same volume as that of the stability domain, namely
\begin{equation}
r_{\alpha,\boldsymbol{\vartheta},N_\mathrm{t}} = \left( \frac{2 A_{\alpha,\boldsymbol{\vartheta},N_\mathrm{t}} }{\pi^2} \right)^{1/4} \, .
\label{radius4D}
\end{equation}

The numerical computation of the DA is equivalent to computing Eq.~\eqref{eqn3o}, and this can be performed by considering $K$ steps in the angle $\alpha$ and $L$ steps in the angles $\vartheta_i$. In this case, the DA reads
\begin{equation}
r_{\alpha,\boldsymbol{\vartheta},N_\mathrm{t}}  = \left[ \frac{\pi}{2 \,K L^2} \sum_{k=1}^{K} \sum_{l_1,l_2=1}^L 
[r(\alpha_k,\boldsymbol{\vartheta}_\mathbf{\ell},N_\mathrm{t})]^4 \sin 2 \alpha_k \right]^{1/4} \quad \text{where} \quad \mathbf{\ell}=(l_1,l_2) \, . 
\nonumber
\end{equation}

The discretisation of the coordinates that are used to define the set of initial conditions whose orbit is evaluated introduces a numerical error in the determination of the DA. The discretisation of the angles $\vartheta_i$ generates a relative error proportional to $L^{-1}$, which corresponds to a trapezoidal integration rule. Note that a better estimate of the error, i.e. a scaling as $L^{-2}$, requires some regularity for the derivative of the function $r(\boldsymbol{\alpha},\boldsymbol{\vartheta},N_\mathrm{t})$. This is probably not the case at the border of the stability domain, and for this reason, the more pessimistic estimate of the error is assumed. The discretisation of the angle $\alpha$ gives a relative error proportional to $K^{-1}$, while the discretisation of the radius $r$ gives a relative error proportional to $J^{-1}$, where $J$ is the number of amplitude steps.
 
From these considerations, one immediately concludes that the integration steps should be optimised to produce comparable errors, i.e. $J \propto K \propto L$. In this way, neglecting the constants that are in front of the error estimates, one obtains a relative error of $1/(4J)$ by evaluating $J^4$ orbits, i.e. $N_\mathrm{t} J^4$ iterates. The fourth power in the number of orbits comes from the dimensionality of the phase space and makes a precise estimate of the dynamic aperture very CPU time consuming. 

In the context of 6D beam dynamics, it should be noted that the earlier arguments can be modified by introducing a coordinate system analogous to polar coordinates in 6D, and one could apply the approach to reduce the 6D sample of the phase space to a faster 3D sample. Nevertheless, this simple extension from the 4D scenario focuses solely on the mathematical aspects, overlooking the inherent physics. Specifically, if $\delta$ represents the relative momentum offset with respect to the nominal momentum value, and $\rho(\delta)$ represents the longitudinal beam distribution, in general, the DA does not go to zero when $\delta$ approaches $\delta_{\rm max}$. Hence, the geometry of the phase-space region in which the orbits are bounded is not spherical in the coordinates $(x,y,\delta)$. Therefore, we better determine $r_{\alpha,\theta,N_\mathrm{t}} (\delta)$ in its parametric dependence on $\delta$ and then define
\begin{equation}
\langle r_{\alpha,\theta,N_\mathrm{t}} (\delta) \rangle_{\delta} = 2 \int_0^{\delta_{\rm max}} r_{\alpha,\theta,N_\mathrm{t}}(\delta) \rho(\delta) \dd\delta \, ,  
\end{equation} 
which correspond to averaging the DA considered as a function of, among the other usual variables, the momentum offset. 

Alternative approaches can be devised to mitigate the unfavourable effects of phase-space dimensionality. In fact, it is possible to perform only a two-dimensional sampling of the phase space by selecting initial conditions of the form $(x_1, 0, x_2, 0)$ and the previous considerations can then be easily adapted~\cite{PhysRevE.53.4067}. This approach is justified if the orbit is far from low-order resonances, as in this case the variables $\vartheta_1, \vartheta_2$, although starting from zero, sample the entire interval of possible values, and the orbit effectively spans the four-dimensional phase space. 

In addition to these refinements in phase-space sampling, DA computation could benefit from parallelised algorithms~\cite{Giovannozzi:317866} to reduce the total time required to perform DA evaluation. Continuing this trend, the volunteer computing platform LHC@home~\cite{Hoimyr:2012,Barranco:2301793} has also been successfully implemented, with an immense benefit to DA studies carried out at CERN in the design stage of the LHC and its luminosity upgrade.

However, the challenges posed by the number of turns employed to simulate orbits in the DA computation persist without any computational method to mitigate them. For instance, a typical duration of the injection process or the physics fill of the LHC is on the order of \SI{30}{minutes} and \SI{12}{hour} respectively, corresponding to approximately \num{2e7} and \num{4.8e8} turns. Both figures go well beyond current computing capabilities. The only successful approach found to date involves attempting to derive a scaling law for the DA in relation to the number of turns. This method allows for the determination of the DA for a realistic, albeit not directly computable, number of turns, from simulations conducted for a manageable, though unrealistically low, number of turns.

Although it is clear that the DA decreases monotonically as a function of $N_\mathrm{t}$, finding a direct solution beyond this basic observation is not straightforward. The solution was obtained by considering the Nekhoroshev theorem providing an estimate of the stability time in a Hamiltonian system~\cite{Nekhoroshev:1977aa}. This theorem is a fundamental one as it provides very generic results (see also Refs.~\cite{Turchetti:1990aa,Bazzani:1990aa} and references therein), which is an essential point as it provides a solid foundation to the model describing the time-evolution of the DA. The initial investigations~\cite{invlog} provided a simple scaling law that was successfully applied to the study of the DA evolution with time. More recently, an in-depth revision of the scaling law~\cite{VanderVeken:2018sqp,Bazzani:2019csk} provided two elegant forms of it, namely
\begin{equation}
\begin{split}
D(N_\mathrm{t}) & =\rho_\ast \left ( \frac{\kappa}{2 \text{e}} \right )^\kappa \, \frac{1}{ \ln^\kappa \frac{N_\mathrm{t}}{N_0}} \, , \\ 
D(N_\mathrm{t}) & = \rho_\ast \displaystyle{\frac{1}{\left[-2 \, \text{e} \, \lambda \,\mathcal{W}_{-1}\!\!\:\!\left(
-\frac{1}{2 \, \text{e} \, \lambda}\left( \frac{\rho_\ast}{6} \right)^{1/\kappa} \, \left( \frac{8}{7} N_\mathrm{t} \right)^{-1/(\lambda \, \kappa)} \right)
		\right]^{\kappa}}} \, , \phantom{\times}
\label{eq:DAmodel}
\end{split}
\end{equation}
where $\rho_\ast, \kappa, N_0, \lambda$ are the parameters to be determined by fitting the numerical data expressing the time evolution of the DA, $\text{e}$ is the basis of the natural logarithm and $\mathcal{W}_{-1}$ is the negative branch of the Lambert function~\cite{corless:1996}. 

The second model reduces to the first one in the limit $\lambda \to 0$ and it is customary to set $N_0=1$, for the first model, or $\lambda=1/2$, for the second model. With these simplifications, both models depend only on two free parameters, namely $\rho_\ast, \kappa$. The Nekhoroshev theorem provides the meaning of these quantities: $\rho_\ast$ is linked to the convergence radius of the perturbative series involved in the proof of the theorem and, from the physical point of view, is linked to the strength of the non-linear effects present in the system under consideration. $\kappa$ is a function of the dimension of the phase space~\cite{Nekhoroshev:1977aa}.

\begin{figure}
    \centering
    \includegraphics[width=0.5\textwidth]{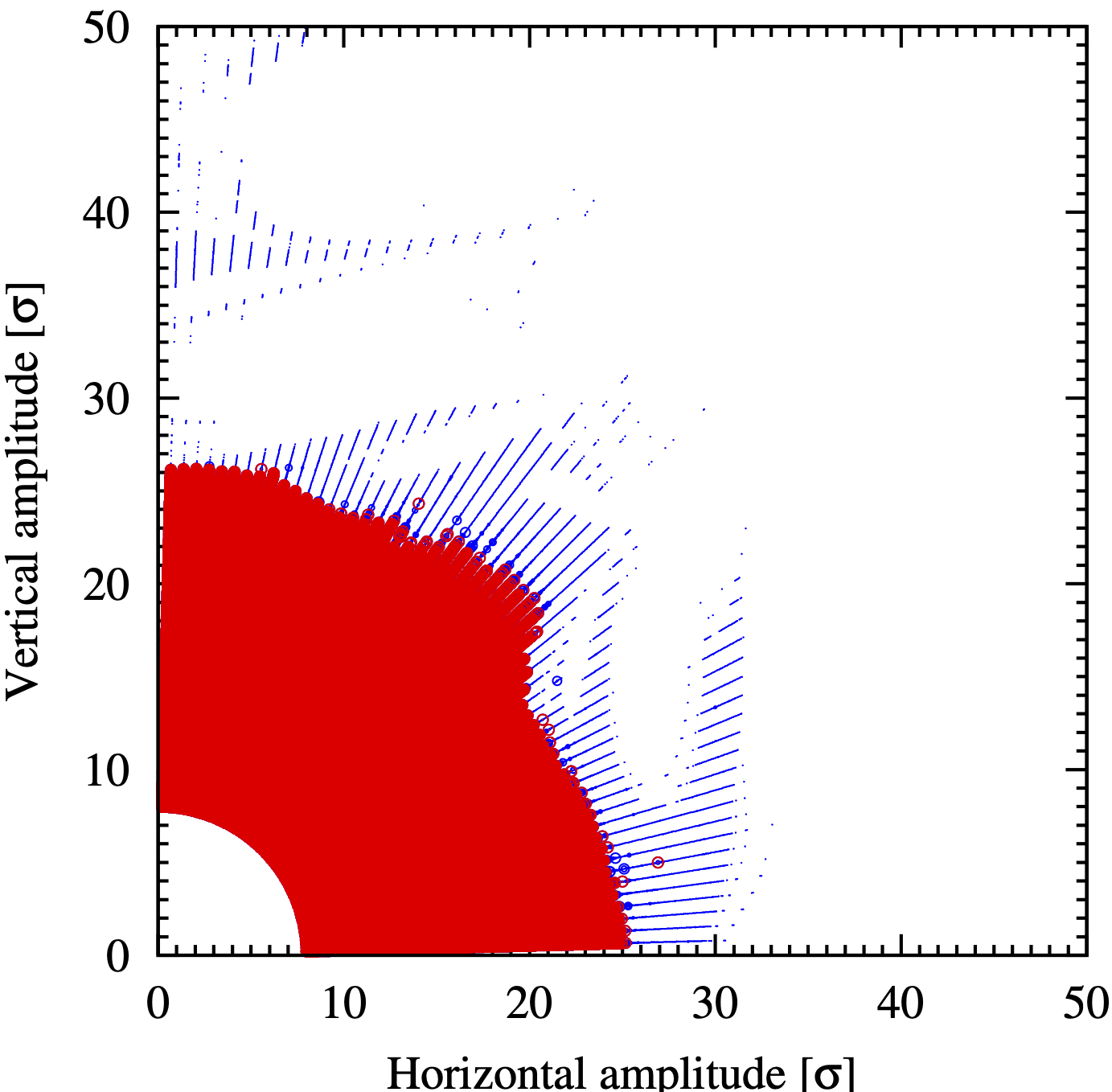}
    \caption{DA of a model of the LHC ring in physical space. The red points indicate initial conditions that are stable up to the maximum number of turns (\num{1e5}) of the numerical simulation. The blue points indicate initial conditions with unbounded orbits and their size is proportional to the number of turns for which their motion is still bounded. The simulation is carried out by performing a scan on a 2D grid, i.e. with initial coordinates of the form $(x_1, 0, x_2, 0)$. (Adapted from Ref.~\cite{da_and_losses}).}
    \label{fig:da-example}
\end{figure}

Figure~\ref{fig:da-example} shows an example of the DA for a model of the LHC ring. The initial conditions selected on a Cartesian grid are of the form $(x_1, 0, x_2, 0)$. We note that only the first quadrant needs to be explored, as the scan is actually performed on the linear invariants in $x_1$ and $x_2$, and the invariants are positive definite functions. The orbit of each initial condition is computed up to \num{1e5} turns, and at each turn, the distance from the origin is evaluated. If it exceeds a prescribed value, the initial condition is labelled unbounded (or unstable, blue points). An initial condition whose distance always remains below the threshold is labelled as bounded (or stable, red point). To speed up numerical simulations, initial conditions close to the origin, which are known \textit{a priori} to be stable, are not tracked, which explains the white region around the origin. The complex geometry of the stable domain is clearly visible. 

\begin{figure}
    \centering
    \includegraphics[trim= 15truemm 0truemm 20truemm 20truemm,width=0.49\textwidth,clip=]{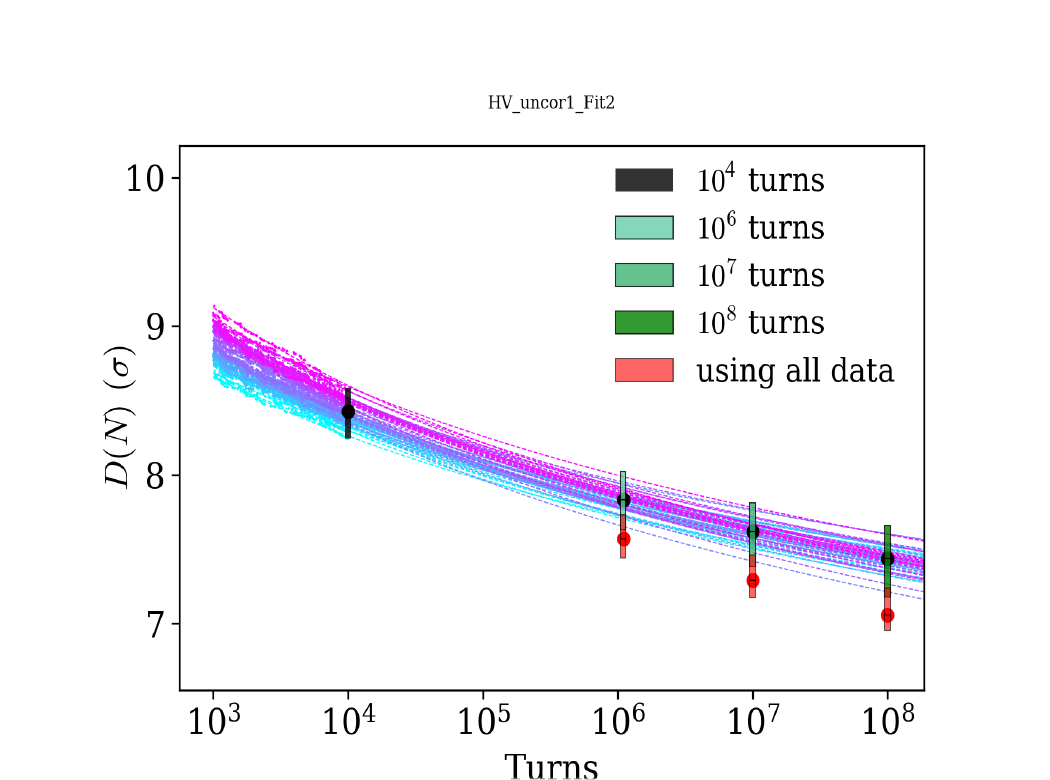}
    \includegraphics[trim= 15truemm 0truemm 20truemm 20truemm,width=0.49\textwidth,clip=]{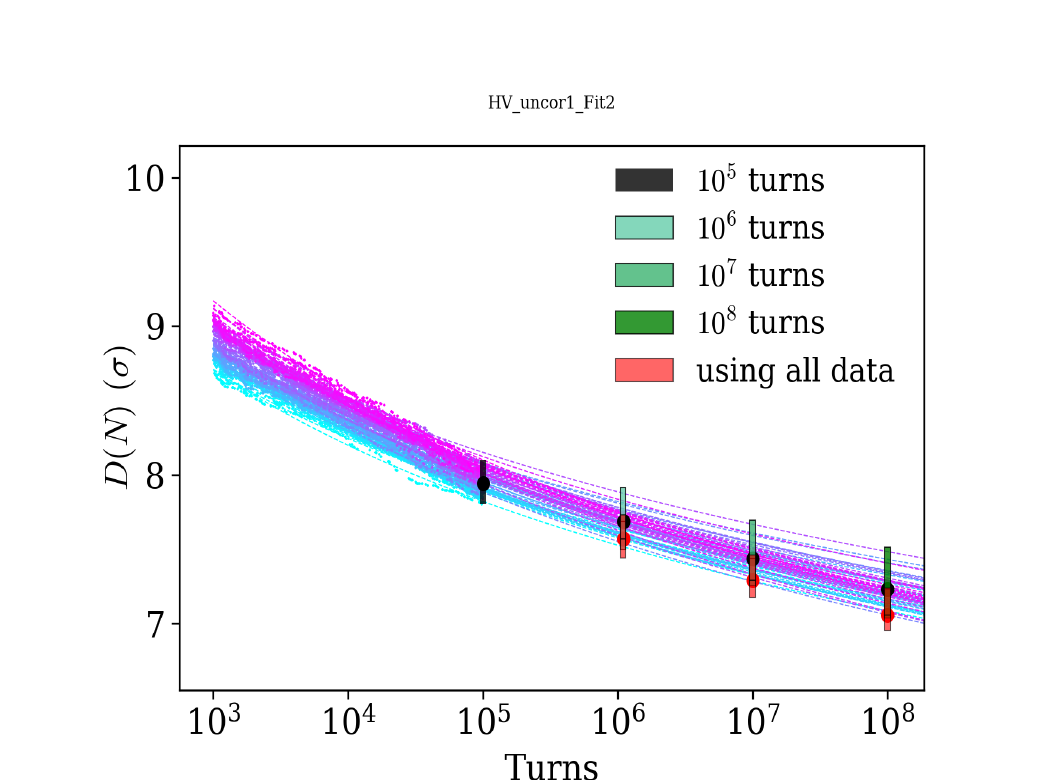}
    \includegraphics[trim= 15truemm 0truemm 20truemm 20truemm,width=0.49\textwidth,clip=]{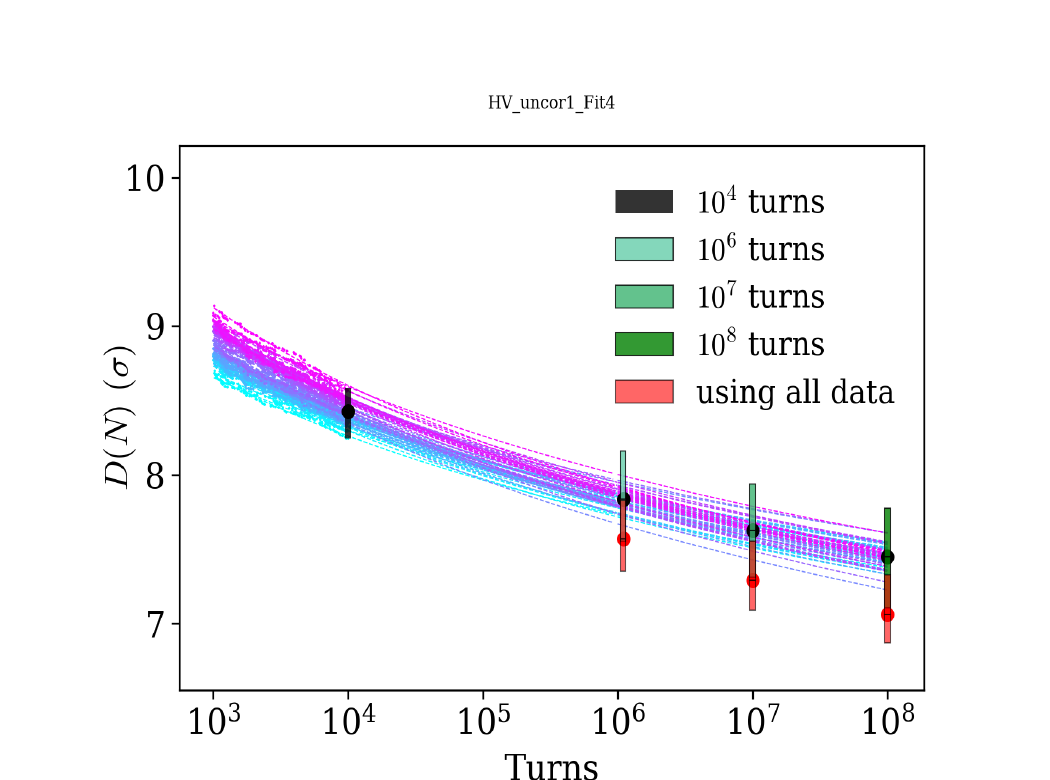}
    \includegraphics[trim= 15truemm 0truemm 20truemm 20truemm,width=0.49\textwidth,clip=]{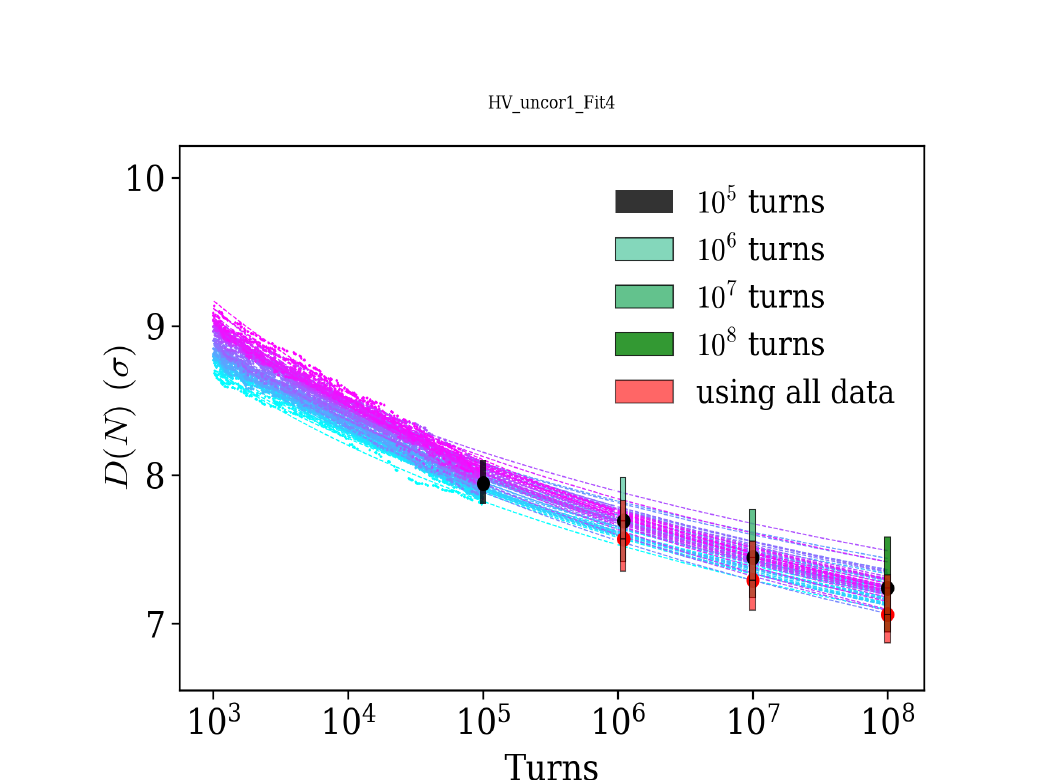}
    \caption{Results of the DA extrapolation obtained from the proposed models in Eq.~\ref{eq:DAmodel} (first row, simplified model, second row, Lambert-function model) built using different number of turns for the DA simulations (\num{1e4}, left, \num{1e5}, right). The numerical simulations are performed using a model of the LHC ring that includes sixty realisations of the non-linear magnetic field errors. The results for all sixty realisations of the LHC ring are plotted, with the numerical data used to build the DA models shown together with the extrapolated curves up to \num{1e8} turns. (Adapted from Ref.~\cite{Bazzani:2019csk}).}
    \label{fig:da-scaling}
\end{figure}

Figure~\ref{fig:da-scaling} reports the analysis of the DA simulations using the models described in Eq.~\ref{eq:DAmodel}. The tracking studies are performed using a model of the LHC that comprises sixty realisations of the non-linear magnetic field errors and computing the orbits up to \num{1e6} turns. For each realisation, the DA is computed using the information from the numerical data, and the DA model is fitted using the same numerical data. However, the fit is performed using a subset of the available data, namely, selecting only the information up to \num{1e4} or \num{1e5} turns (left and right columns, respectively). The DA models are then used to predict the DA for \num{1e6}, \num{1e7}, and \num{1e8} (the last two values are beyond computing capabilities). The predictive power of the models is compared with the DA of numerical simulations up to \num{1e6} turns, and the agreement is excellent, of the order of 2\%-4\%. The data presented in the first row are based on the simplified DA model, whereas those presented in the second row are based on the Lambert-function model. These plots serve as outstanding illustrations of the effectiveness of the method based on DA scaling laws.

The last important step is the link between the DA and the evolution of the beam intensity in a storage ring or collider under the influence of non-linear effects. Models describing the evolution of the DA as a function of time are essential in this regard. It is assumed that if the orbit of an initial condition crosses the region of bounded motion up to turn $N_1$, then it is lost at turn $N_1+1$. With this in hand~\cite{da_and_losses,titze:2021}, one can proceed by assuming Gaussian distributions for the transverse degrees of freedom
\begin{equation}
\rho(q,p; \epsilon) = \frac{1}{2 \pi \epsilon} \exp\left(- \frac{q^2}{2 \epsilon} - \frac{p^2}{2 \epsilon} \right)
\label{eq:hat_rho}
\end{equation}
where $\epsilon$ stands for the beam emittance and the actual beam distribution reads
\begin{equation}
f(x_1,p_{x_1}, x_2, p_{x_2}) = N_\mathrm{p} \rho(x_1,p_{x_1}; \epsilon_{x_1}) \rho(x_2, p_{x_2}; \epsilon_{x_2}) \, ,
\end{equation}
where $N_\mathrm{p}$ is the total beam intensity. 

Using $x_i^2 + p_{x_i}^2 = 2 J_{x_i} , i=1,2$, $f$ can be expressed in action-angle variables:
\begin{equation}
f(J_{x_1}, J_{x_2}, \varphi_{x_1}, \varphi_{x_2}) = \frac{N_\mathrm{p}}{4 \pi^2 \epsilon_{x_1} \epsilon_{x_2}} \exp\left(-\frac{J_{x_1}}{\epsilon_{x_1}}
- \frac{J_{x_2}}{\epsilon_{x_2}}\right)\, ,
\end{equation}
and 
\begin{equation}
\hat f(J_{x_1}, J_{x_2}) = \int_0^{2 \pi} \int_0^{2 \pi} f(J_{x_1}, J_{x_2}, \varphi_{x_1}, \varphi_{x_2}) \dd \varphi_{x_1} \dd \varphi_{x_2} = \frac{N_\mathrm{p}}{\epsilon_{x_1} \epsilon_{x_2}} \exp\left(-\frac{J_{x_1}}{\epsilon_{x_1}}
- \frac{J_{x_2}}{\epsilon_{x_2}}\right)\, .
\end{equation}

We introduce the coordinates $r \in [0, \infty ]$ and $\theta \in [0, \pi/2]$ as follows:
\begin{subequations}
\begin{align}
\sqrt{J_{x_1}} = \sqrt{\epsilon_{x_1}} r \cos \theta \, , \\
\sqrt{J_{x_2}} = \sqrt{\epsilon_{x_2}} r \sin \theta \, ,
\end{align}
\end{subequations}
and we define the following quantity
\begin{equation}
r_\mathrm{max}(\theta)=\frac{R \sqrt{\epsilon_{x_1} \epsilon_{x_2}}}{\sqrt{\epsilon_{x_2}\cos^2\theta + \epsilon_{x_1} \sin^2\theta}} \qquad R \in [0, \infty[ \, .
\label{eq:rmax}
\end{equation}

We are interested in the fraction of particles contained within a specific region given by $r \leq r_\mathrm{max}(\theta) , \theta \in [0, \pi/2 ] $. This corresponds to computing the surviving particles $S$, i.e.\ those particles that are located inside the DA, which is given by
\begin{equation}
S(R) = \int_0^\infty \int_0^\infty \Theta_R(r, \theta) \hat f(J_{x_1}, J_{x_2}) \dd J_{x_1} \dd J_{x_2} , \qquad 
\Theta_R(r, \theta) = \left\{\begin{array}{cl} 1 & \text{for $r \leq r_\mathrm{max}(\theta)$} \\ 0 & \text{for $r > r_\mathrm{max}(\theta)$ \, ,} \end{array} \right .
\label{eq:rd}
\end{equation}
which becomes
\begin{equation}
\begin{split} 
S(R) & = 4 N_\mathrm{p} \int_0^{\pi/2} \dd\theta \int_0^{r_\mathrm{max}(\theta)}  \dd r\, \mathrm{e}^{-r^2} r^3 \cos \theta \sin \theta \\
& =4N_\text{p} \int_0^{\pi/2}\dd\theta\,\cos\theta\sin\theta \int_0^{r_\mathrm{max}(\theta)}\dd r\, r^3 e^{-r^2} \\ 
&= 2N_\text{p} \int_0^{\pi/2} \dd\theta\,\sin\theta\cos\theta \qty[1-(1+r^2_\mathrm{max}(\theta))e^{-r^2_\mathrm{max}(\theta)}] \\
& = N_\text{p}\qty[1 - \frac{\epsilon_{x_1} e^{-R^2 \epsilon_{x_2}} - \epsilon_{x_2} e^{-R^2 \epsilon_{x_1}}}{\epsilon_{x_1} - \epsilon_{x_2}} ] \, .
\end{split}
\end{equation}

All these relationships represent the number of charged particles in the region $r \le r_\mathrm{max}(\theta)$ and $R=D(N)$, where $D(N)$ is the DA value that varies with the number of turns $N$. Hence, one obtains
\begin{equation}
S(N) = N_\mathrm{p} \qty[1 - \frac{\epsilon_{x_1} e^{-D(N)^2 \epsilon_{x_2}} - \epsilon_{x_2} e^{-D(N)^2 \epsilon_{x_1}}}{\epsilon_{x_2} - \epsilon_{x_1}} ] \, ,
\label{eq:losses1}
\end{equation}
which represents a relationship between the evolution of the DA and the number of surviving particles $S(N)$. This is especially significant because it creates a connection between an abstract observable, the DA, and a measurable quantity, the beam intensity. Additionally, this connection links the beam lifetime to the DA, offering another highly useful relationship. This is crucial because it allows the DA concept to transition from being confined to tracking simulations and design studies to becoming relevant in accelerator performance and optimisation. As a result of these considerations, a method has been proposed to measure DA in a circular accelerator~\cite{PhysRevAccelBeams.22.034002}, circumventing the need to displace beams at high amplitudes. This is a major benefit for high-energy superconducting rings, in which the DA exceeds the limits of deflection devices and in which sudden losses that could lead to a magnet quench must be strictly avoided. These relationships have also been used to develop models that characterise the evolution of instantaneous luminosity when accounting for burn-off and non-linear effects~\cite{Giovannozzi:2018wmm,Giovannozzi:2018igq,amezza}.

The drawback of this method lies in its lack of self-consistency, as it overlooks any processes through which the beam distribution may change over time. Within this framework, only the DA changes, while the beam distribution is presumed to remain constant. To appropriately predict changes in the beam distribution and use these changes to evaluate beam losses or other relevant beam parameters, diffusion models should be employed.
\subsection{Diffusive motion} \label{sec:diff}
The emergence of a finite DA\footnote{For a linear system, the DA is infinite.} results from non-linear effects in beam dynamics. In a quasi-integrable Hamiltonian system, the phase space is divided into invariant tori interspersed with chaotic regions. A stronger perturbation of the original integrable system leads to more extensive chaotic layers. In extreme scenarios, some or all initial chaotic layers can become linked, as explored in the work of Chirikov~\cite{chirikov:1979} (see also Ref.~\cite{lichtenberg:1992}). When phase space has large weakly-chaotic regions, the dynamics can mimic a diffusion process, allowing the application of related concepts to examine how the beam distribution evolves.

The conditions described are readily observable in the motion of charged particles within a circular accelerator. This is not solely due to the presence of non-linear field errors but also owing to time-dependent effects connected to ripples in the power supplies of the ring magnets. Although these influences are initially periodic over time with diverse frequency spectra, collectively they result in a comprehensive stochastic behaviour. Consequently, it is reasonable to assume that the beam dynamics in the weakly-chaotic region is governed by a quasi-integrable Hamiltonian system subject to stochastic perturbations (refer to the appendix of Ref.~\cite{bazzani2020diffusion}). Associated with this is a Fokker-Planck (FP) equation~\cite{Bazzani9948}, which describes the average evolution of the beam distribution and includes absorbing boundary conditions. This mathematical model aligns well with the physical reality, as modern superconducting accelerator rings are equipped with collimators, which can be viewed as absorbing boundary conditions in the context of beam dynamics.

Although these arguments serve mainly to justify the well-established use of the diffusive method for non-linear beam dynamics, the new approach introduced in Refs.~\cite{Bazzani:2019lse,bazzani2020diffusion,montanari:ipac2021:tupab233,our_paper9,montanari:ipac22-mopost043,montanari:ipac2023-wepa022,Montanari:2863942,montanari:2025} makes a significant advancement by proposing a specific form for the diffusion coefficient and connecting it to the perturbative series used in the context of the Nekhoroshev theorem~\cite{Nekhoroshev:1977aa}. This innovative proposal creates a link between the DA and the interpretation of beam dynamics via a FP equation, and its implications have yet to be thoroughly investigated.

Considering a diffusion framework for the evolution of the distribution of initial conditions in the action variable $I$, in the case of a one-dimensional scenario (refer to the appendices of Ref.~\cite{bazzani2020diffusion} for mathematical insights), the FP equation is applicable and is given by
\begin{equation}
\frac{\partial \rho}{\partial t} = 
\frac{\varepsilon^2}{2}\frac{\partial}{\partial I}\mathcal{D}(I)
\frac{\partial}{\partial I}\rho(I,t) \, ,
\label{fokker2}
\end{equation}
where $\varepsilon$ serves as a scaling parameter related to the intensity of the non-linear disturbance in the quasi-integrable Hamiltonian system. According to the prediction of the Nekhoroshev theorem, the action-diffusion coefficient can be expressed as follows:
\begin{equation}
\mathcal{D}(I) = c \, \exp\left [-2 \left (\frac{I_\ast}{I}\right )^{1/(2\kappa)}\right ] \, .
\label{diffnek}
\end{equation}
This expression is considered a reasonable assumption for describing action diffusion across extensive areas of weakly-chaotic phase space. The constant $c$ is determined by normalising the diffusion coefficient through the equation
\begin{equation}
c \, \int_0^{I_{\rm abs}}  \exp\left [-2 \left (\frac{I_\ast}{I}\right )^{1/(2\kappa)}\right ] \dd I = 1 \, ,
\label{normal}
\end{equation}
where $I_{\rm abs}$ denotes the location of the absorbing boundary condition. The parameters $(\varepsilon, \kappa, I_\ast)$ defining the diffusion model~\eqref{fokker2} and~\eqref{diffnek} can be interpreted using Nekhoroshev's theorem: $\varepsilon$ is a dimensionless parameter that quantifies the non-linear effects on the beam, resulting in a rescaling of time; $\kappa$ is derived from the perturbative series' analytic structure and is mainly influenced by the dimensionality of the phase space and the nature of the non-linear terms in the series, but not their strength. It is also connected to the corresponding parameter that defines the DA scaling laws. Meanwhile, $I_\ast$ pertains to the intensity of the non-linear terms. In particular, $\varepsilon$ and $I_\ast$ are theoretically interconnected, since a change in the action scale can alter the intensity of the perturbation. This correlation is disrupted by the boundary condition, as the absorbing barrier remains invariant to $\varepsilon$'s global scaling yet varies with scaling of the action.

\begin{figure}[htb]
\centering
\includegraphics[trim= 20mm 60mm 40mm 55mm, width=0.59\linewidth,clip=]{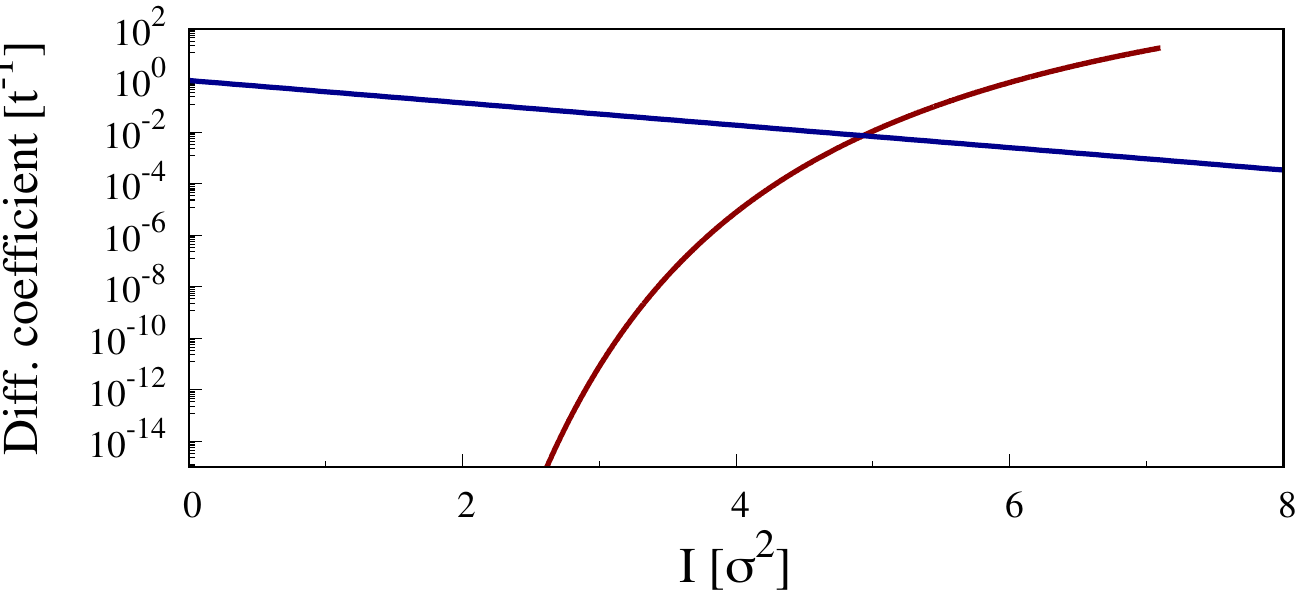}
\includegraphics[trim= 20mm 60mm 40mm 55mm, width=0.59\linewidth,clip=]{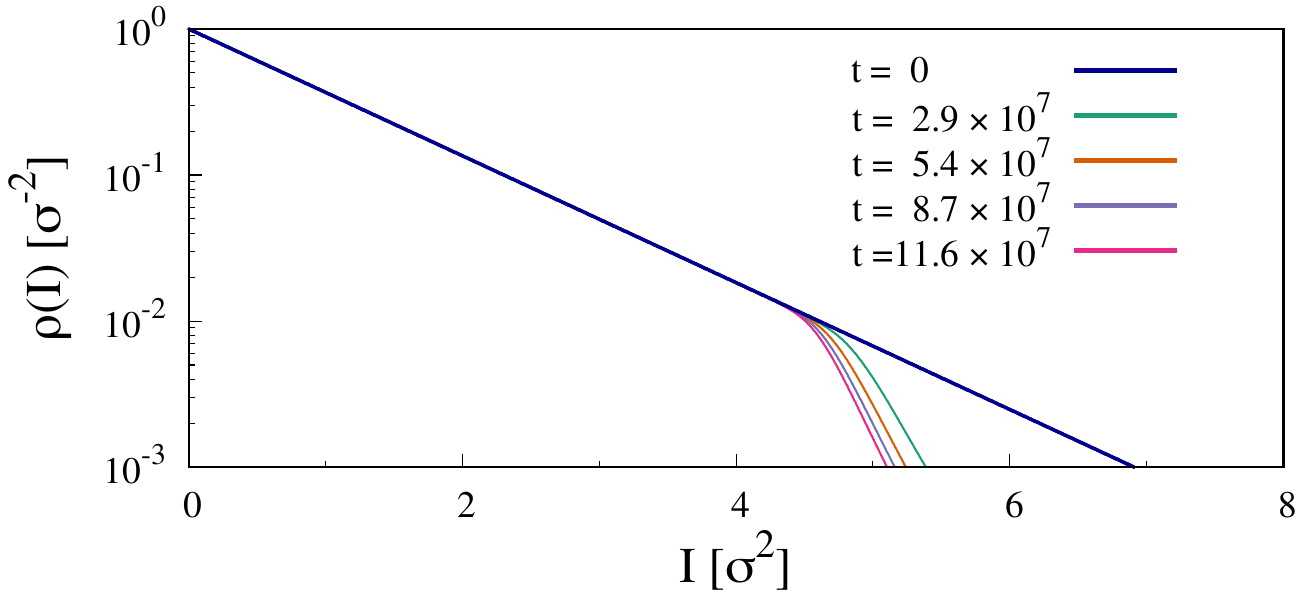}
\caption{Top: the behaviour of the Nekhoroshev diffusion coefficient~\eqref{diffnek} (depicted by the red curve) is presented as a function of $I$ for the parameters $\kappa=0.33$ and $I_\ast=21.5$. To emphasise the characteristics of the function~\eqref{diffnek}, the exponential curve $\exp(-I)$ (blue curve) is also included. Bottom: the evolution of a one-dimensional Gaussian distribution is presented using the Fokker-Planck equation~\eqref{fokker2} incorporating the Nekhoroshev diffusion coefficient shown in the upper plot. The action variable and $I_\ast$ are expressed in units of the sigma of the distribution of the initial conditions (Adapted from Ref.~\cite{bazzani2020diffusion}.)}
\label{fpfig}
\end{figure}

Figure~\ref{fpfig} (top) illustrates the behaviour of the diffusion coefficient as expressed in Eq.~\eqref{diffnek} (marked by the red curve), using the parameter values $\kappa$ and $I_\ast$ from beam measurement campaigns. For comparison, an exponential function is represented by the blue curve. The characteristics of the Nekhoroshev functional form are evident, with a rapid exponential decline as $I \to 0$ and a trend toward saturation as $\mathcal{D}(I) \to 1$ when $I \to \infty$. This limit has only mathematical significance because the action is initially bounded by the DA and subsequently by the dimensions of the vacuum chamber and collimators. The lower plot of the figure illustrates the evolution of an initial Gaussian distribution\footnote{Note that in terms of physical variables, a Gaussian distribution corresponds to an exponential distribution in the action variable.}. The way in which the diffusion coefficient behaves is evident in how the initial distribution evolves. Over time, modifications occur at a decreasing rate, primarily impacting the distribution's tails while leaving the inner region unchanged.

The functional form of $\mathcal{D}(I)$ qualitatively aligns with the requirements to account for previously observed phenomena (refer to, e.g., the discussion in Ref.~\cite{gerasimov1992applicability}), such as rapid changes of the diffusion coefficient with respect to the action variable. Additionally, the mathematical principles underlying the Nekhoroshev theorem lend strong support to this hypothesis. Consequently, this methodology appears to be highly suitable for depicting the evolution of beam distribution tails. 

The experimental exploration of the suggested diffusive model remains a vibrant area of research. These beam measurements pose significant challenges due to the tiny nature of the beam losses being assessed. When conducted in superconducting rings, it is necessary to strictly manage these losses to prevent magnet quenches. Measurement campaigns have been executed at CERN's LHC employing two complementary methodologies. In the first approach, the preparatory phase of the experiment involves enlarging the beam emittance to expedite diffusion. Intensity decay is tracked over time, revealing that losses occur predominantly near a primary collimator, which serves as an absorbing boundary condition in the diffusion analysis~\cite{bazzani2020diffusion}. The results of these measurements accurately validated the intensity decay pattern, which was closely replicated using numerical simulations with the FP equation~\eqref{fokker2}, where the free parameters of the diffusion coefficient were aligned with the observed loss data.

The second method relies on the collimator system as a key tool in investigating the diffusion process. Essentially, the collimators probe the beam distribution by gradually scraping small portions of it as the collimator jaw moves. Although this technique is not new, it had to be modified to accommodate measurements of the proposed form of $\mathcal{D}(I)$, due to the highly non-linear dependency on free parameters~\cite{our_paper9,Montanari:2863942}. Unlike the first method, which depends on measuring the total beam intensity, this approach focuses on precisely measuring the tiny losses caused by the collimator jaws that scrape particles along their movement path, posing a significantly greater challenge for researchers~\cite{montanari:ipac2021:tupab233,montanari:ipac22-mopost043,montanari:ipac2023-wepa022}. However, successful experiments were conducted and data analysis revealed that the measured diffusion agrees well with the proposed model~\cite{montanari:2025}. 

The successful results of the recent diffusion measurements are an essential step in the promotion of the proposed diffusion framework. However, a crucial element yet to be incorporated into this model is the behaviour of the distribution's core. In practice, real accelerators frequently exhibit emittance growth, typically arising from random dipolar excitations. Clearly, the diffusion coefficient \eqref{diffnek} alone is insufficient to induce growth in the inner region of the distribution. This challenge could be addressed by modifying the Nekhoroshev form of the diffusion coefficient with an added constant term, which would contribute to both the inner distribution growth and the tail evolution. These considerations are currently being actively investigated.
\section{Exploiting non-linear beam dynamics} \label{sec:expl}
\subsection{Dynamic use of stable islands} \label{sec:dyna-isl}
The effective use of the possibilities presented by non-linear effects began with combining stable resonance islands with adjusting the accelerator's betatronic tune to confine particles within these islands. This form of beam manipulation is a straightforward application of adiabatic theory pertinent to separatrix crossing~~\cite{neish1975,NEISHTADT198158,NEISHTADT1986,an4,NEISHTADT1991,an6,an9,an10,Neishtadt2013,Neishtadt_2019}. It is important to note that within a Hamiltonian system, fixed points can be classified as either elliptic (stable) or hyperbolic (unstable). These names not only describe the geometry of curves near the fixed point, but also reflect the dynamics or stability of the fixed point. For completeness, there is a third category, the parabolic fixed point, though it is a degenerate case and is not pertinent to our discussion. A separatrix is a Hamiltonian curve that intersects a hyperbolic fixed point. Motion near the separatrix slows exponentially, disrupting any adiabatic condition and necessitating a specific theory to understand separatrix crossing. It is clear that in the absence of explicit time dependency in a Hamiltonian system, a separatrix acts as a boundary for orbits: orbits originating outside the separatrix remain outside, while those starting inside the separatrix stay within it.
\begin{figure}
\centering
    \includegraphics[trim=45truemm 185truemm 45truemm 45truemm,width=0.8\columnwidth,clip=]{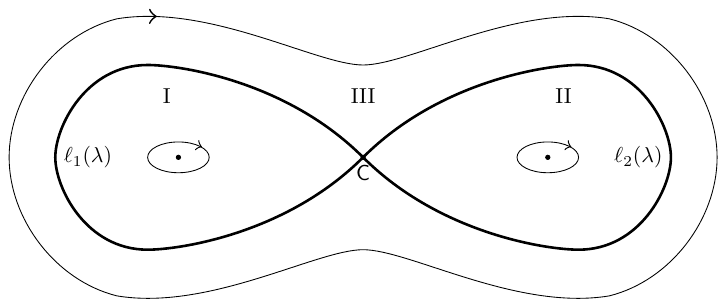}
    \caption{A generic phase-space $(q,p)$ portrait divided into three regions ($\mathrm{I}$, $\mathrm{II}$, $\mathrm{III}$) by separatrices $\ell_1(\lambda)$ and $\ell_2(\lambda)$. (Adapted from Ref.~\cite{our_paper4}).}
    \label{fig:genericphsp}
\end{figure}

Consider a Hamiltonian $\ham(p, q, \lambda = \epsilon \, t)$ where $\epsilon \ll 1$ indicates that the parameter $\lambda$ changes slowly over time. The phase space, depicted in Fig.~\ref{fig:genericphsp}, shows the separatrix passing through the point $C$, which divides the space into three distinct regions, each containing one trajectory. The main conclusion of separatrix crossing theory is that if the change in $\lambda$ is adiabatic, the transition process through the separatrix is probabilistic. For an initial condition starting in Region~$\mathrm{III}$, the likelihood of transition to either Region~$\mathrm{I}$ or $\mathrm{II}$ is expressed by~\cite{neish1975}
\begin{equation} 
  \mathcal{P}_{\mathrm{III} \to \mathrm{I}} = \frac{\Theta_\mathrm{I}}{\Theta_\mathrm{I} + \Theta_\mathrm{II}} \, \qquad \mathcal{P}_{\mathrm{III} \to \mathrm{II}} = 1 - \mathcal{P}_{\mathrm{III} \to \mathrm{I}} \, ,
  \label{eq:neish}
\end{equation}
where
\begin{equation}
 \Theta_i = \dv{A_i}{\lambda}\eval_{\tilde\lambda} = \oint_{\partial A_i} \pdv{\ham}{\lambda} \eval_{\tilde\lambda} \, \dd t \qquad i = \mathrm{I}, \, \mathrm{II} \, ,
 \label{eq:prob}
\end{equation}
with $A_i$ being the area of the region $i$, $\partial A_i$ representing the boundary of the region $i$, and $\tilde\lambda$ the value of $\lambda$ at the point of crossing the separatrix. It is important to note that if $\mathcal{P}_{\mathrm{III} \to i} < 0$, then $\mathcal{P}_{\mathrm{III} \to i}$ is adjusted to zero; if $\mathcal{P}_{\mathrm{III} \to i} > 1$, it is set to one, ensuring that $\mathcal{P}$ remains a legitimate probability function. This implies that to harness the adiabatic trapping phenomena effectively, one must be able to calculate the surfaces of different phase-space regions and choose appropriate system parameters to manage these surfaces. Consequently, this facilitates the modification of the probability function, allowing for comprehensive control over the capture process. We emphasise that the theory has been thoroughly developed for Hamiltonian systems with a single degree of freedom. However, extending this to systems with two or more degrees of freedom remains incomplete, primarily because of geometric challenges. In parallel, the theory pertaining to maps or discrete-time systems is yet to be fully formulated, with only partial efforts thus far.

\begin{figure*}
\centering
    \includegraphics[trim=0truemm 0truemm 0truemm 0truemm,width=0.6\textwidth,clip=]{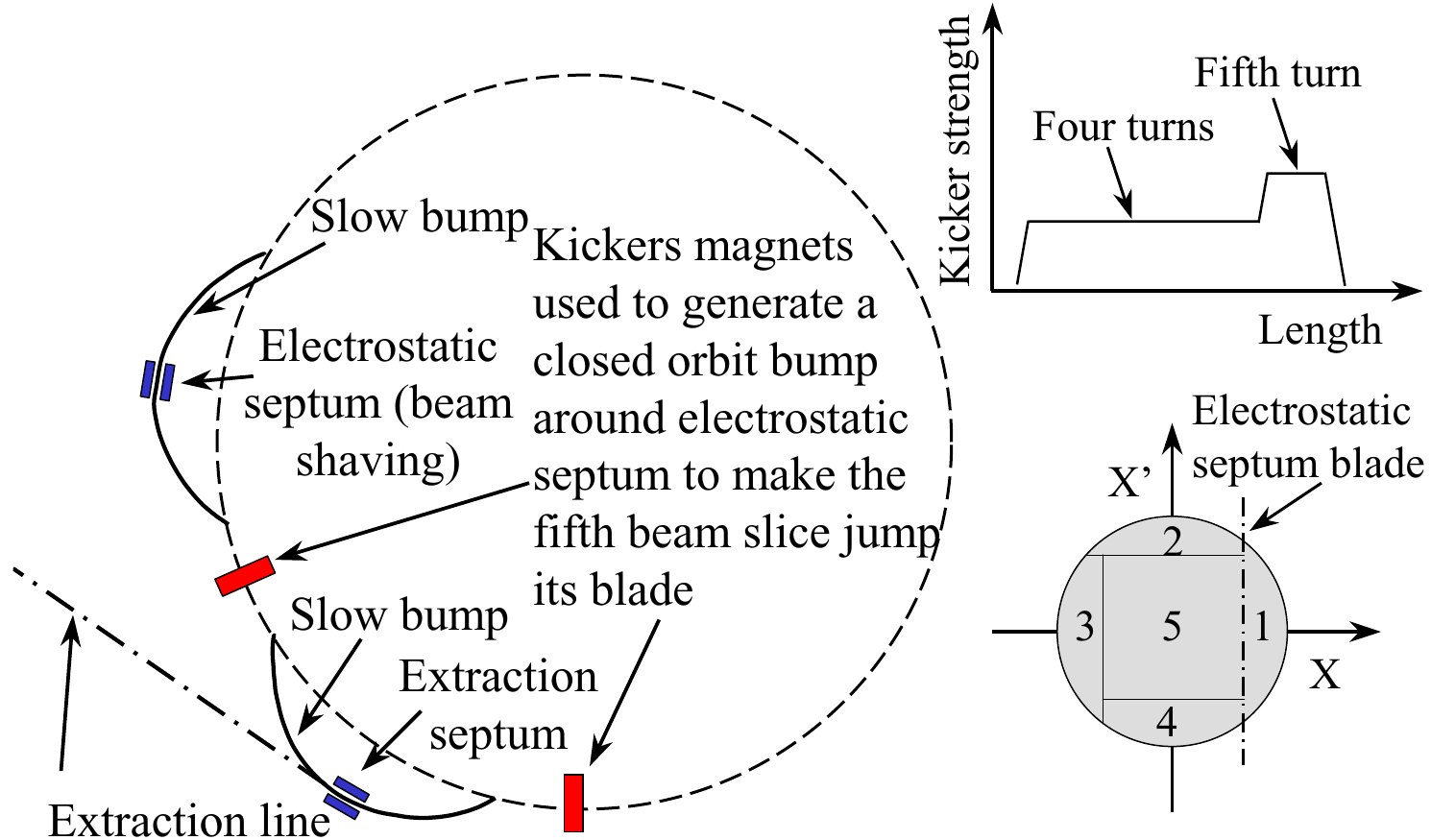}
\caption{Overview of the CT extraction technique. The intricate network of slow bumps employed to alter the closed orbit is illustrated, aiming to direct the beam towards the slicing device and the main extraction septum. The horizontal plane shaving technique relies on a tune value of $6.25$ and involves deflecting part of the beam after it moves through the foil of an electrostatic septum. A diagram of the time-dependent strength variation of the fast dipoles used to push the beam through the slicing device is also provided. (Adapted from Ref.~\cite{PhysRevLett.88.104801}).}
\label{fig:CTsketch}
\end {figure*}

This mathematical framework has been used and applied to find a solution to a very practical issue in the management of the high-intensity proton beams used at the CERN Super Proton Synchrotron (SPS) for the fixed-target physics programme. These challenging beams were extracted from the Proton Synchrotron (PS) using a complicated and intrinsically lossy extraction method, the so-called Continuous Transfer (CT)~\cite{bovet73}. The goal was to devise an extraction method to ensure the fastest filling of the SPS with the high proton flux required for fixed-target experiments. The optimal solution was based on the extraction of the beam over five PS turns using the principle described in Fig.~\ref{fig:CTsketch}. 
Two gradual shifts, known as closed orbit bumps, are generated to guide the beam closer to the slicing device and the extraction septum. Concurrently, the horizontal tune is adjusted to $6.25$, causing the beam to rotate \SI{90}{\degree} in the horizontal phase space with each revolution around the ring. The initial stage of the extraction process involves activating fast-deflecting dipoles, or kickers, to create a closed bump around the slicing device and force the beam across it. The portion of the beam within the electrostatic septum, which acts as the slicing device, is deflected towards the extraction septum to proceed into the transfer line leading to the SPS. The horizontal tune facilitates dividing the beam into five segments, the last of which is extracted using an extra deflection kicker.

\begin{figure}[htb]
\centering
{\includegraphics[width=0.510\linewidth,clip=]{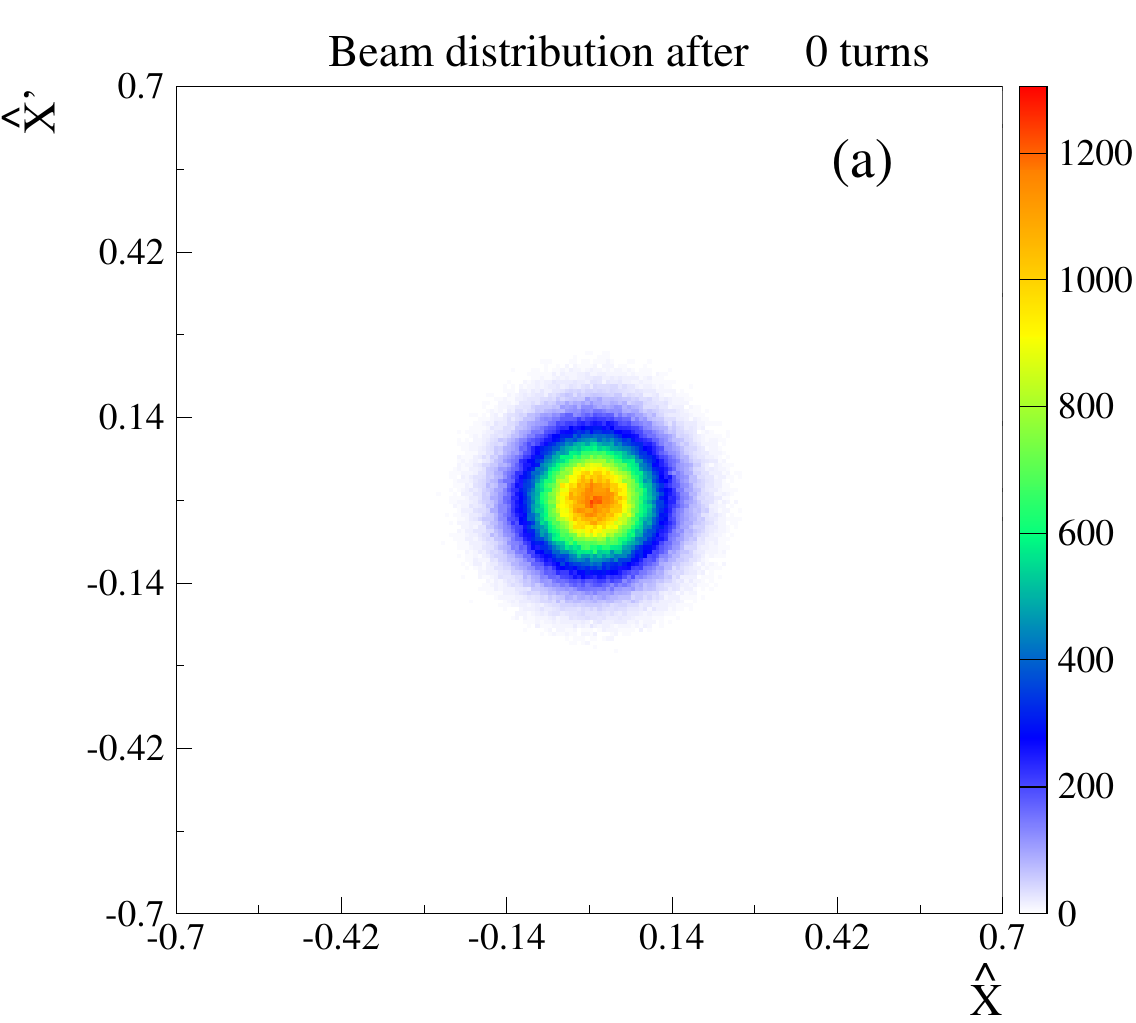}}
{\includegraphics[width=0.454\linewidth,clip=]{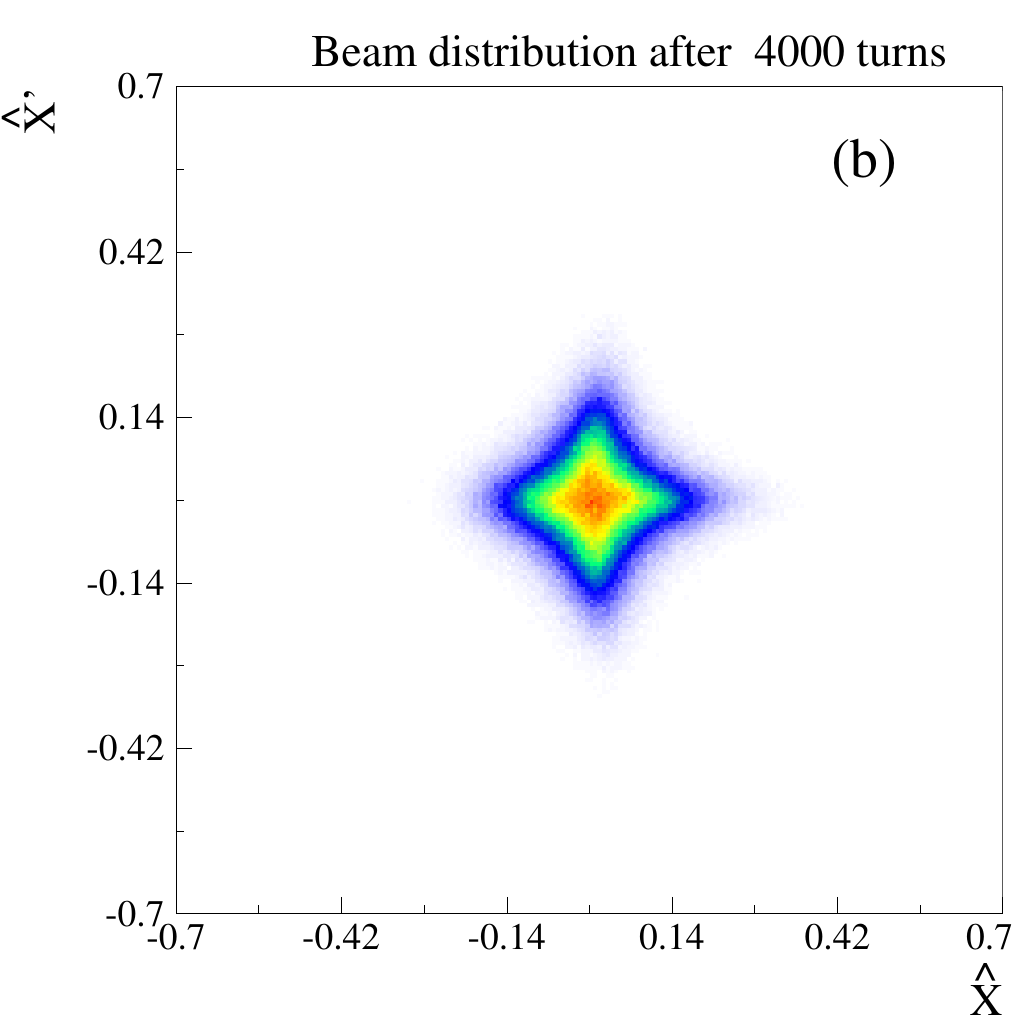}} \\
{\includegraphics[width=0.510\linewidth,clip=]{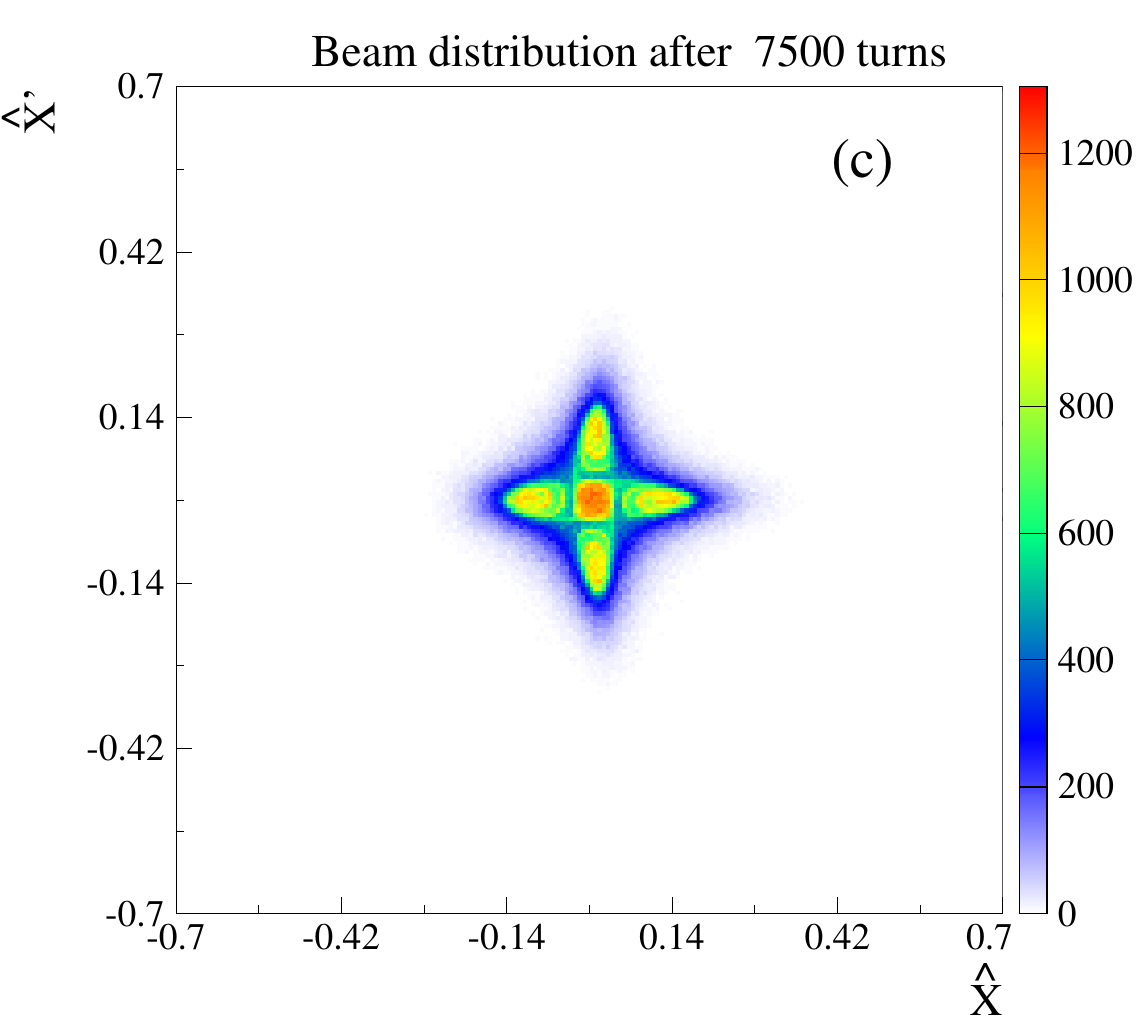}}
{\includegraphics[width=0.454\linewidth,clip=]{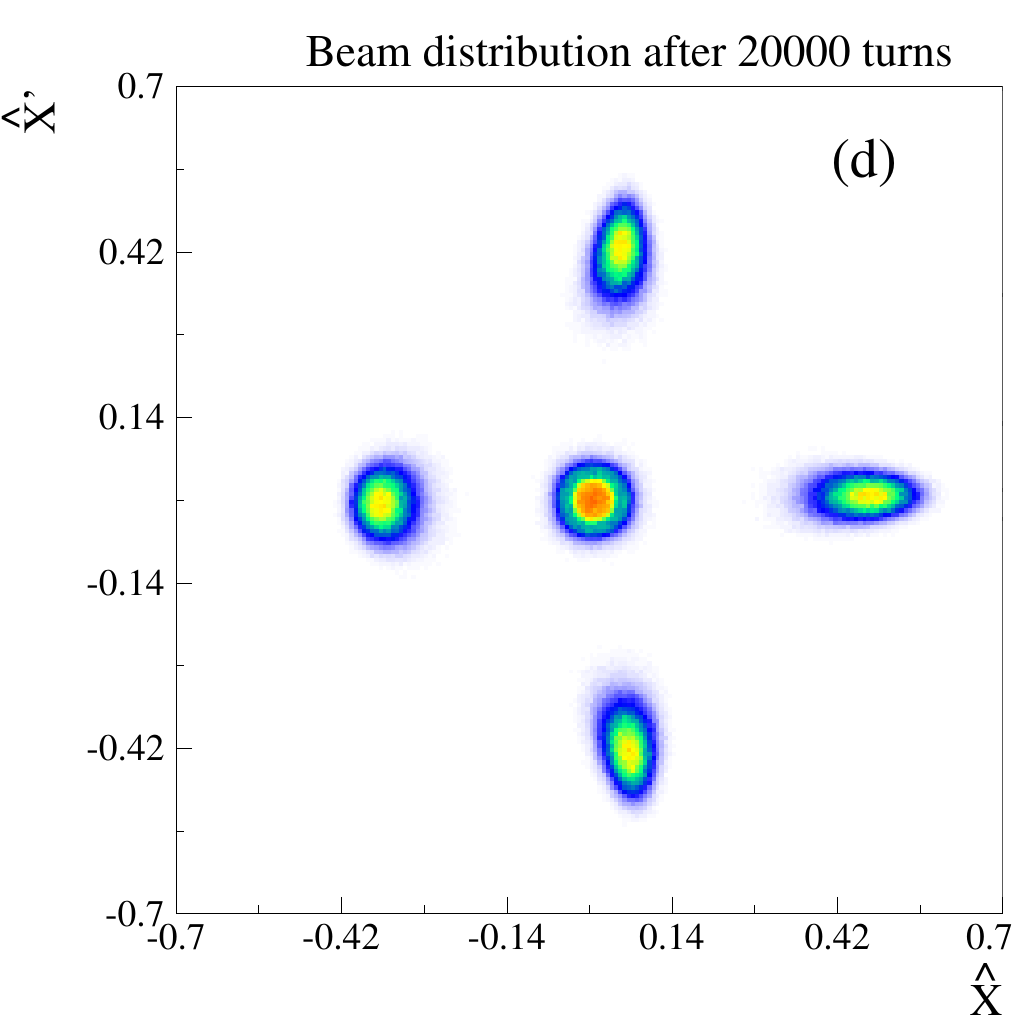}}  \\
\caption{During the trapping process that involves crossing a fourth-order resonance, the beam distribution evolves in a notable manner. Initially, it is represented by a Gaussian centred at zero; ultimately, it transforms into five distinct Gaussians, one remaining at the centre and four corresponding to the positions of the islands. This outcome results from adiabatic trapping, obtained by varying the horizontal tune through the resonant value of $6.25$, aided by phase-space transport that facilitates the movement of the beamlets trapped in the islands towards higher amplitudes. (Adapted from Ref.~\cite{PhysRevSTAB.7.024001}).}
\label{fig:res4}
\end{figure}

Unavoidable beam losses are generated by the beam-foil interaction, and scattered particles spread over approximately a third of the PS ring downstream of the electrostatic septum irradiating the devices in their path~\cite{Gilardoni:EPAC08-THPC047,PhysRevSTAB.14.030101}. Non-linear beam dynamics allows for the division of the beam without using a physical device, utilising adiabatic trapping. Specifically, by generating stable islands with sextupole and octupole magnets and varying the horizontal tune of the PS ring, particles can be confined within these regions, thereby dividing the beam into several beamlets: one located at the phase space's centre and others within the islands. This operation is fundamental to MTE, serving as a substitute for CT extraction~\cite{PhysRevLett.88.104801}. Figure~\ref{fig:res4} illustrates this essential procedure, showing the changes in the beam distribution as the horizontal tune is modified. Initially, a Gaussian distribution in the centre splits into five Gaussian distributions as the tune passes the $6.25$ mark. Although the resonant tune value remains identical for both CT and MTE, this is the only feature they share, a reflection of their fundamentally different natures.  

It is crucial to emphasise that the final five beamlets constitute a structure divided into two segments: the central beam, which shares the same periodicity as the ring's circumference; and the four beamlets within the stable islands forming a unique structure that completes a loop after four revolutions around the accelerator's circumference. This implies that the structure extends beyond the physical length of the accelerator, enabling its extraction over the course of five turns. First, the beamlets, over a total of four turns, then the beam at the centre on a single turn.

The Introduction has succinctly recapped the extensive journey bridging the initial proof-of-principle sketch~\cite{PhysRevLett.88.104801} to the operational use of MTE~\cite{Borburgh:2137954}, marking the conclusion of a demanding phase of formidable challenges.

Concrete applications can frequently inspire new ideas that are broader in scope than the specific problem that prompted them. For example, adiabatic trapping and transport can be employed to devise methods that allow the extraction of beams over varying numbers of turns by choosing a resonance other than the conventional fourth-order~\cite{PhysRevSTAB.7.024001,PhysRevSTAB.12.024003}. Using the principle of time reversal, one can also develop Multi-turn Injection (MTI) strategies~\cite{PhysRevSTAB.10.034001}. This all comes from departing from conventional linear beam dynamics to explore the vast potential offered by non-linear beam dynamics.

Additional steps toward a full exploitation of non-linear beam dynamics were made by considering new Hamiltonian systems that include oscillating magnetic elements and by studying the characteristics of a process that crosses a two-dimensional resonance, i.e. a process that involves both the horizontal and the vertical tunes. 

Feedback systems used to control instabilities in circular accelerators and to excite the beam for the purpose of measuring the ring optics are based on dipole magnets that may also generate oscillating fields. A device that generates oscillating fields creates a dynamics that has previously been examined in the mathematical literature, without considering its physical implications, leading to separatrix crossing phenomena~\cite{Neishtadt2013}. A notable feature of these systems is that the resonance condition is established between the frequency of the unperturbed system and the frequency of the oscillating device. This has significant implications and intriguing side effects. Notably, the tune of the accelerator ring does not need to be altered to achieve separatrix crossing; it can remain constant, with the frequency of the oscillating component changing over time. This could be an important advantage for many applications.  

\begin{figure*}
\centering
    \includegraphics[trim=2truemm 1truemm 4truemm 2truemm,width=.325\textwidth,clip=]{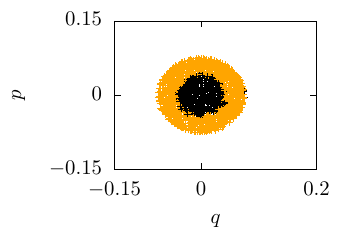}
    \includegraphics[trim=2truemm 1truemm 4truemm 2truemm,width=.325\textwidth,clip=]{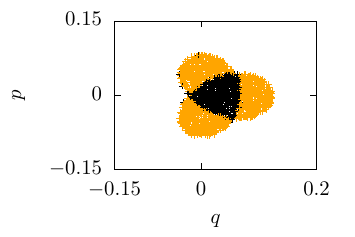}
    \includegraphics[trim=2truemm 1truemm 4truemm 2truemm,width=.325\textwidth,clip=]{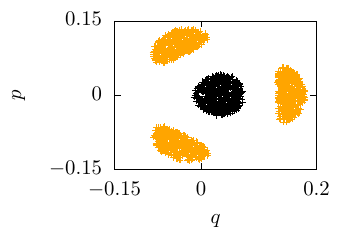}
\caption{The evolution of a collection of particles in phase space is influenced by an oscillating dipole that fulfils a third-order resonance condition with the ring tune. The dipole frequency changes over time moving from left to right. The colours indicate whether the initial condition leads to the central structure or trapping in the islands. (Adapted from Ref.~\cite{our_paper4}).}
\label{fig:splitACdip}
\end {figure*}

Figure~\ref{fig:splitACdip} illustrates the splitting process initiated by an oscillating dipole. The system is subjected to a third-order resonance condition between the ring tune with the oscillating dipole's frequency. Adjusting the dipole's amplitude and frequency causes the initial distribution to divide into four beamlets. Although the resulting beam configuration does not reveal the specific splitting process applied, the fundamental underlying mechanism remains the same: controlled separatrix crossing.

The notion of an oscillating dipole can be naturally extended, at least within a mathematical framework, to encompass higher-order magnets like oscillating sextupoles and octupoles. This extension allows for the development of sophisticated beam manipulation techniques, which aim to achieve either the reduction of the beam emittance~\cite{PhysRevAccelBeams.26.024001} or the cleaning of beam halo, a critical and challenging aspect for optimising the performance of high-energy superconducting accelerators~\cite{capoani:2025}.

\begin{figure*}
\centering
    \includegraphics[trim=0truemm 0truemm 0truemm 0truemm,width=.5\textwidth,clip=]{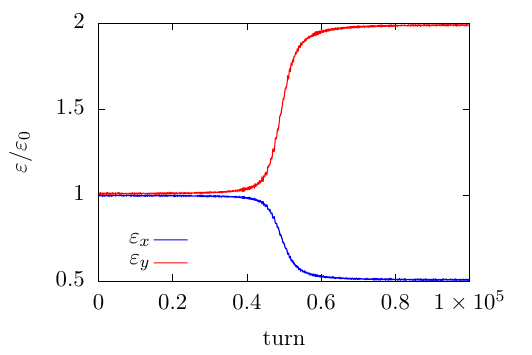}
\caption{The evolution of the horizontal and vertical emittances, normalised to their initial values, during the process of crossing the two-dimensional resonance $Q_x - 2 Q_y= p$, with $p \in \mathbb{Z}$ for the case of a simplified Hamiltonian model. The redistribution of the emittances is clearly visible. (Adapted from Ref.~\cite{our_paper7}).}
\label{fig:Esharing}
\end {figure*}

In this line of investigation, crossing a two-dimensional resonance results in a redistribution of the transverse emittances~\cite{Capoani:2863096,our_paper7,PhysRevAccelBeams.24.094002,PhysRevAccelBeams.25.104001}. Denoting the horizontal and vertical tunes of the ring by $Q_x$ and $Q_y$, a one-dimensional resonance is represented by the relation $n Q_z = p$, where $z=x, y$ and $n, p \in \mathbb{N}$. In contrast, a two-dimensional resonance condition involves both $Q_x$ and $Q_y$ and takes the form $n Q_x + m Q_y= p$, with $n, m, p \in \mathbb{Z}$. When such a two-dimensional resonance is crossed, the emittances are redistributed so that the ratio of the final emittance to the initial one is $\vert m/n \vert$ or $\vert n/m \vert$, depending upon the plane. This indicates that the product of the emittances remains constant throughout the crossing. The redistribution effect is clearly illustrated in Fig.~\ref{fig:Esharing}, which depicts the ratio in the case of crossing a resonance characterised by $n=1$ and $m=-2$ as a function of time during the resonance-crossing process for the case of a simplified Hamiltonian model describing the resonance-crossing process.

These insights into non-linear beam manipulations demonstrate a significant capability for managing transverse emittances, surpassing the strict conservation framework imposed by straightforward linear dynamics. Moreover, there are undoubtedly numerous additional applications that could be developed, beyond those discussed here and explored in recent studies.
\subsection{Static use of stable islands} \label{sec:stat-isl}
Non-linear effects hold significant potential as they facilitate the creation of unique phase-space configurations, such as stable islands and associated separatrices. By thoughtful modification of certain system parameters combined with such phase-space topology, new opportunities can be unlocked. Consequently, trapping and transport of particles is possible, which can be used in a controlled way as elaborated in the preceding section.

The unique topology of such phase-space configurations has the potential for manipulating beam dynamics even in the absence of time dependence, and hence of trapping and transport phenomena. The key observation is that the stable fixed points present in stable islands represent a closed orbit that can be used to accommodate particles. The accelerator ring therefore inherits further closed orbits in addition to the standard closed orbit at the origin of the phase space\footnote{We assume that there are no dipole errors that might alter the standard closed orbit and shift it away from the phase space origin.}. The dynamics around these new closed orbits differs from that around the standard closed orbit, which means that there is the possibility of designing a ring with multiple closed orbits and independent optical parameters. The choice of the period of the stable islands is yet another parameter that can be used to differentiate the motion between the two classes of closed orbit. It should be noted that while the optics of the standard closed orbit is controlled by the quadrupoles, that of the secondary closed orbit is affected by higher-order magnets, because of feed-down effects, and these magnets have no impact on the optics of the standard closed orbit. This opens up the possibility of an independent control of the properties of the two classes of closed orbits. 

\begin{figure}
    \centering
    \includegraphics[trim= 10truemm 35truemm 40truemm 20truemm,width=0.49\textwidth,clip=]{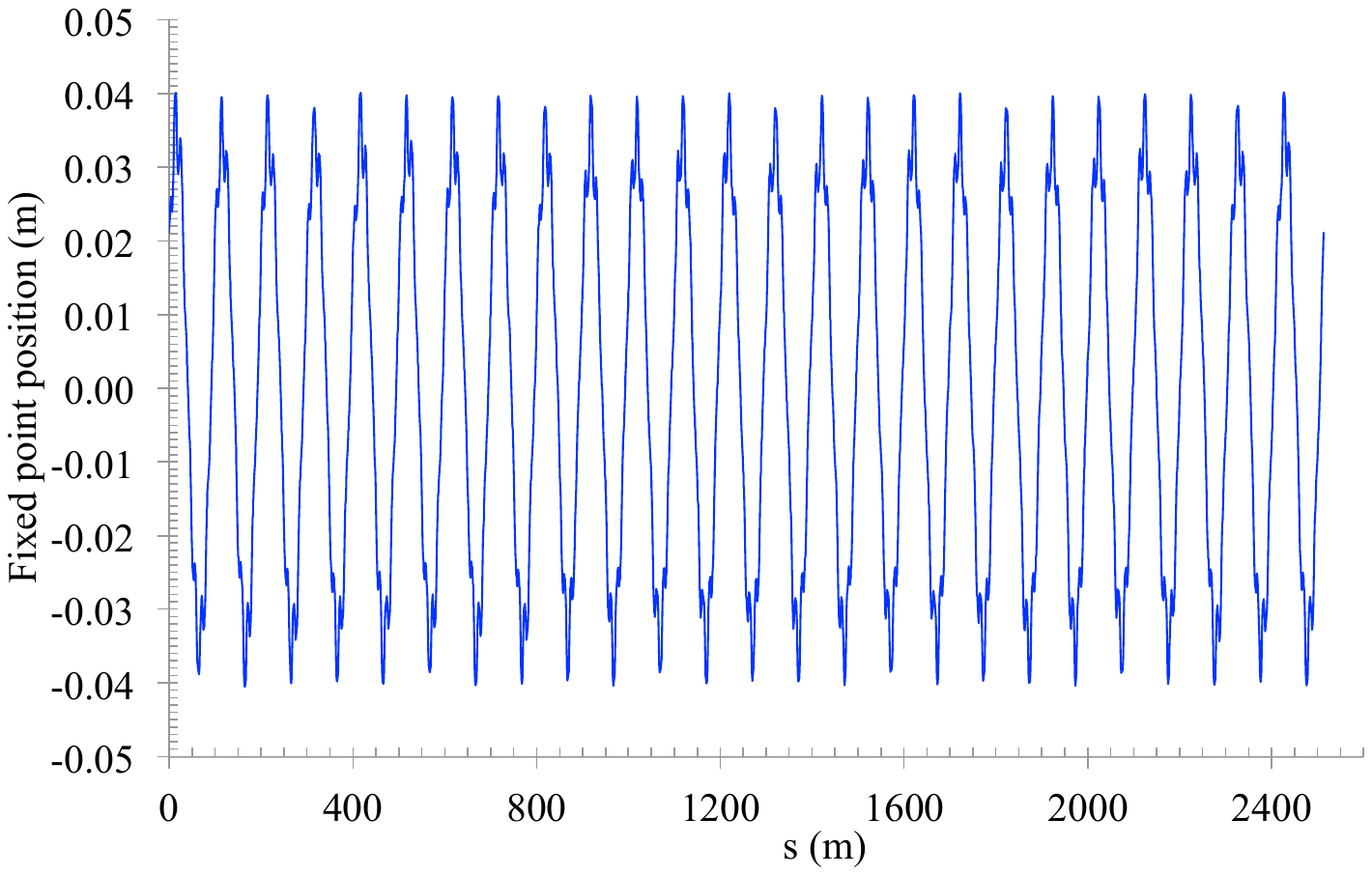}
    \includegraphics[trim= -6truemm 3truemm 10truemm 0truemm,width=0.49\textwidth,clip=]{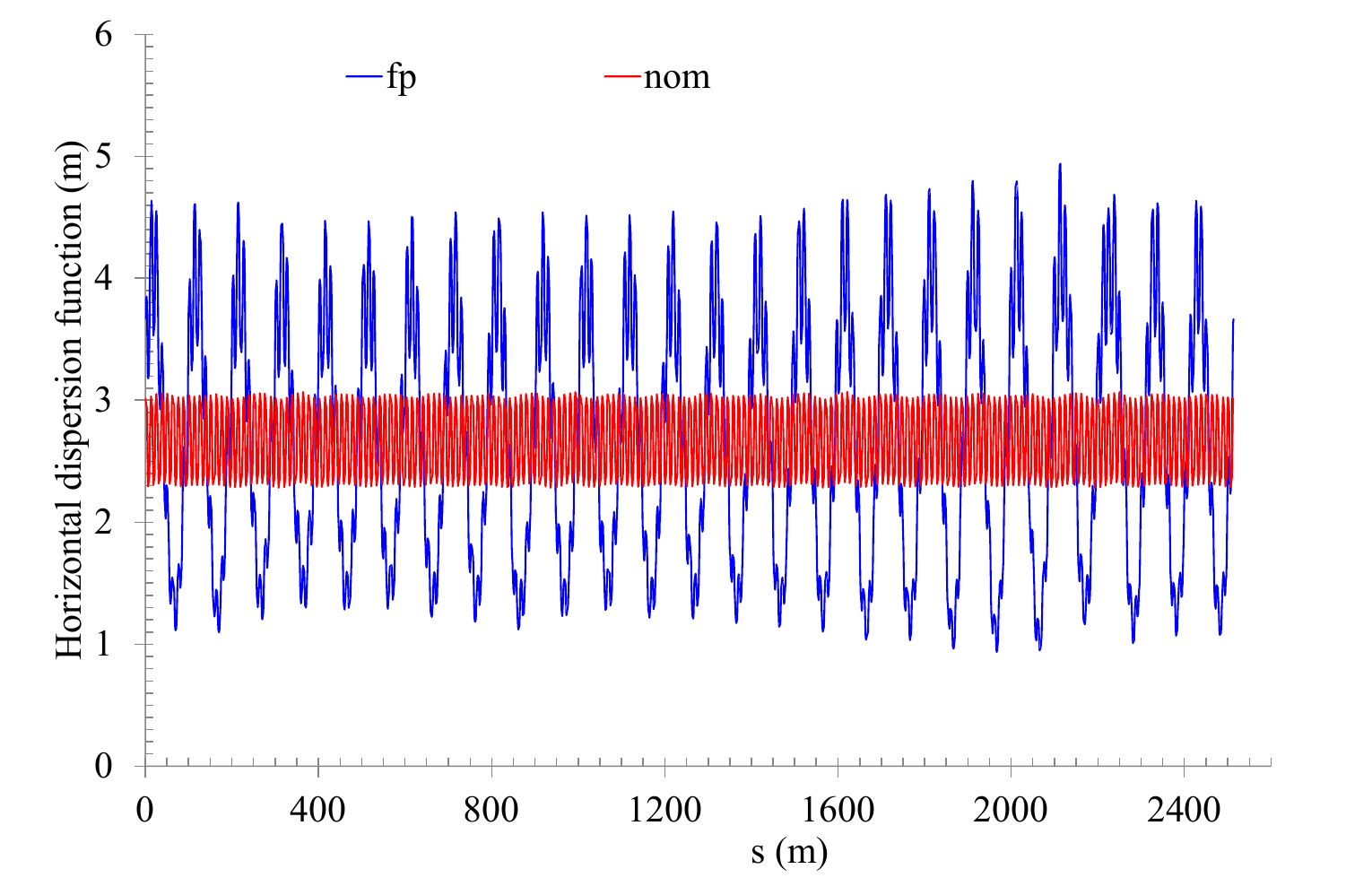}
    \includegraphics[trim= -6truemm 10truemm 10truemm 0truemm,width=0.49\textwidth,clip=]{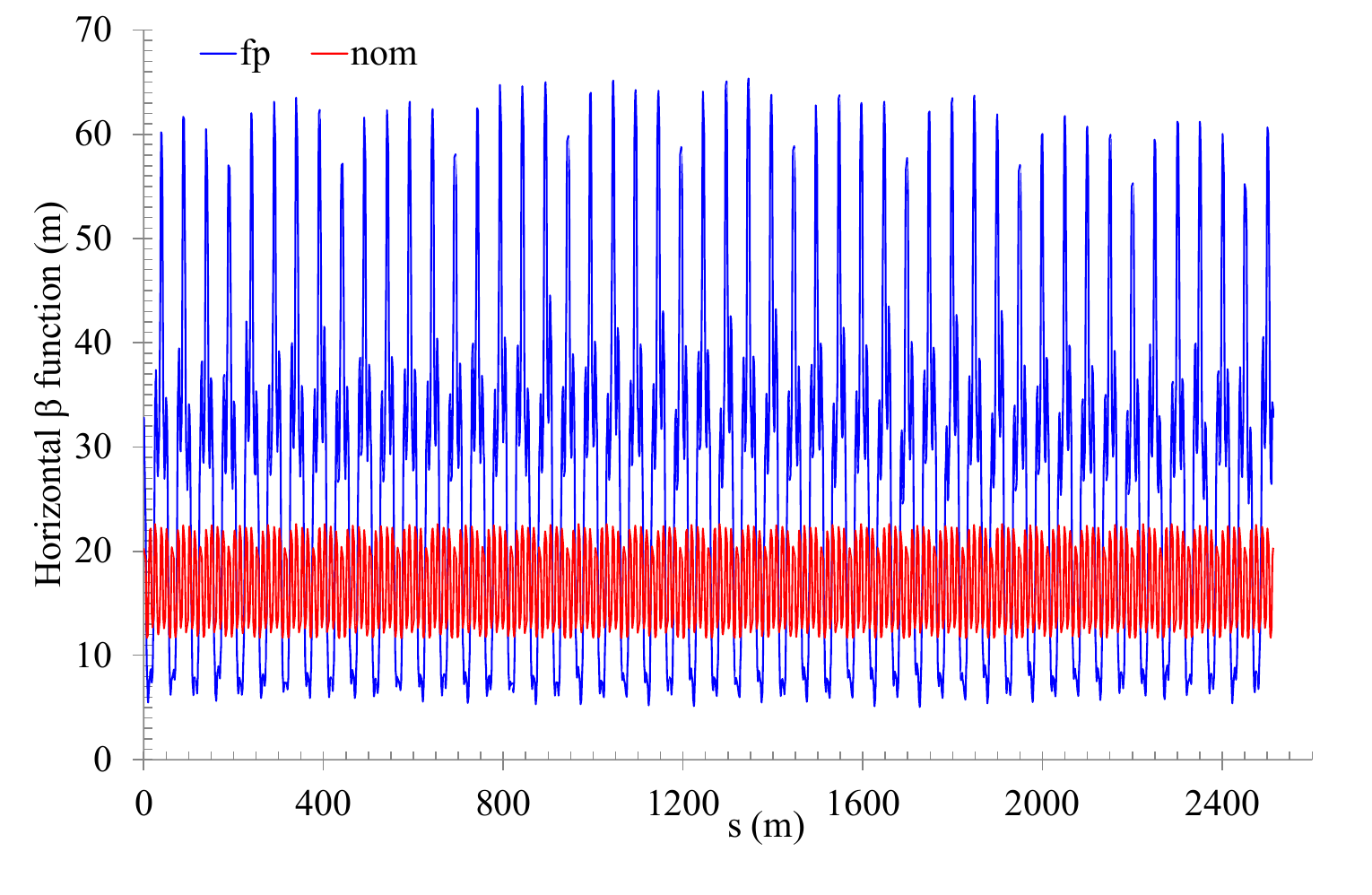}
    \includegraphics[trim= -6truemm 10truemm 10truemm 0truemm,width=0.49\textwidth,clip=]{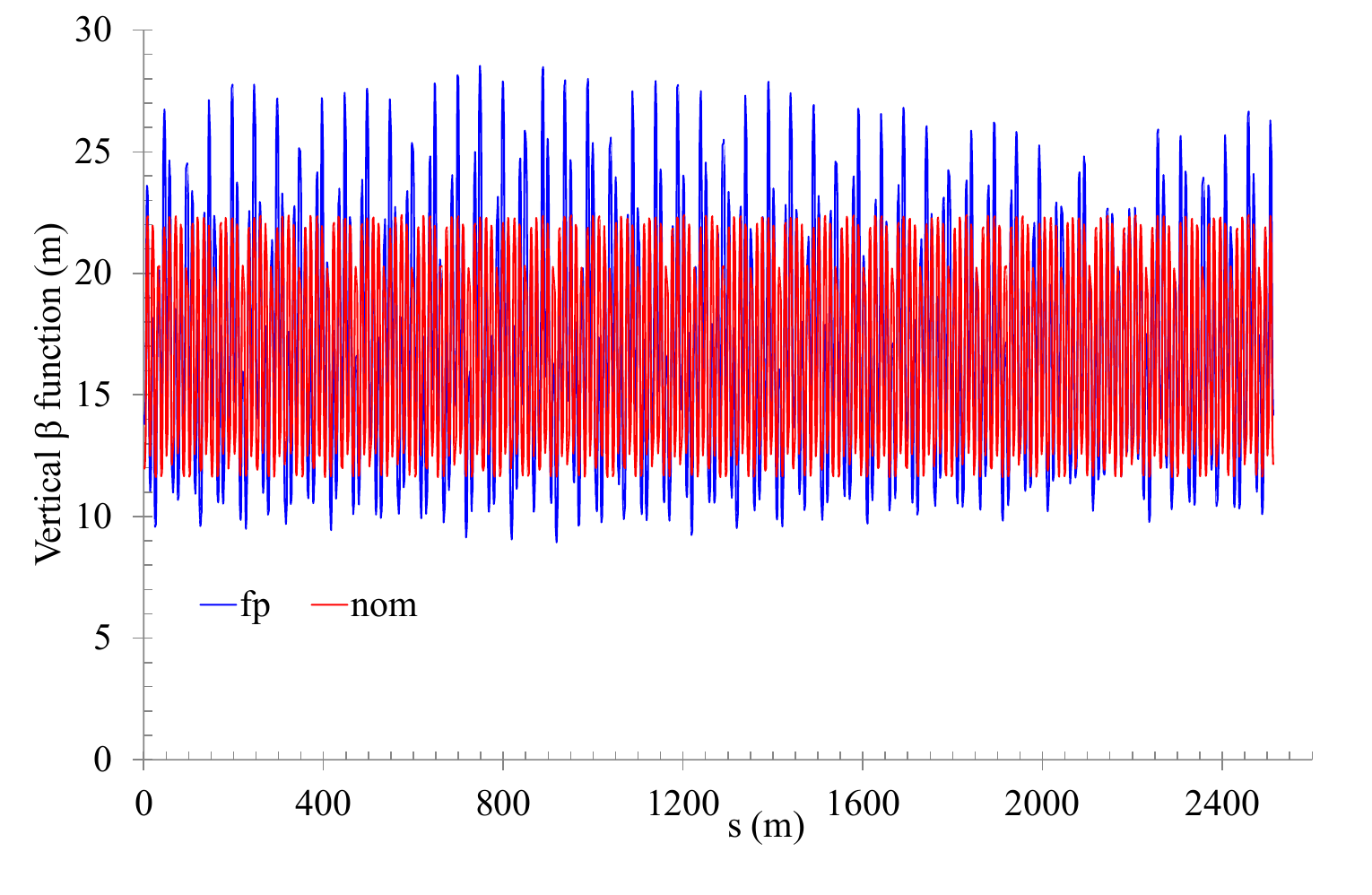}
    \caption{Example of secondary orbit (upper left), dispersion (upper right), and beta-function (lower row) around the circumference of a model of the CERN PS ring. The model considered is actually made of four copies of the PS to satisfy the periodicity of the fixed points of the fourth-order resonance. The red curves represent the optical parameters and dispersion for the standard closed orbit, whereas the blue curves represent the same parameters but for the secondary closed orbit. The difference is clearly visible. (Adapted from Ref.~\cite{GiovannozziEPJPlusTransition}).}
    \label{fig:PS-optics}
\end{figure}

Figure~\ref{fig:PS-optics} illustrates a configuration using a CERN PS ring model, where four instances of the ring have been concatenated to maintain the periodicity of the stable islands and their corresponding fixed points, as this specific case employs the fourth-order resonance. The upper-left graph depicts how the fixed-point position changes along the circumference, while the upper-right graph displays the dispersion, and the lower-row graphs show the horizontal and vertical beta-functions. The red curves correspond to the optical parameters around the standard closed orbit, and the blue curves correspond to the optical parameters around the secondary closed orbit (the fixed point). The stark contrast between the two families of curves highlights the potential of the approach to create different optical conditions inside the same ring. 

A first application of this configuration with multiple closed orbits could be the possibility of designing a fast extraction\footnote{By applying a time reversal, the same argument applies to injection.}, i.e. an extraction of the beam in a single turn, without the use of septum magnets~\cite{PhysRevSTAB.18.074001}. Instead, a kicker is used to deflect the beam and place it in the stable islands. Once there, the beam follows the secondary closed orbit, which can be designed to guide the beam to the extraction channel. 

Another proposal is for non-adiabatic gamma-jump gymnastics to cross the transition energy, $\gamma_\mathrm{t} m_0 c^2$, where $\gamma_\mathrm{t}$ is the relativistic gamma-transition value, $m_0$ the rest mass of the proton, and $c$ the speed of light ~\cite{GiovannozziEPJPlusTransition}. This method was evaluated on a theoretical setup of the CERN PS ring and yielded promising outcomes.

Transition crossing is a delicate moment during the energy ramp in storage rings, in particular when high-intensity beams are used (see, e.g. Refs.~\cite{bryant_johnsen_1993,Lee:2651939} for a general overview of the physics of transition-energy crossing). To avoid or overcome the issues related to the crossing of the transition energy, the ideal solution would be to execute the crossing as quickly as possible to minimise any detrimental effects. Yet, it is often impractical because of the constraints posed by the main magnets and the radio-frequency system. Hence, an effective approach is to adjust the optical parameters of the rings, modifying the dispersion function to quickly alter the transition energy value~\cite{Risselada:213345}. This technique, which is based on the use of families of quadrupoles that are pulsed to create the changes in optical parameters of the entire ring, promotes a faster crossing without directly influencing the beam acceleration process.

The new method takes advantage of the ability to have distinct optical properties for two closed orbits, allowing different values of $\gamma_\mathrm{t}$ for the standard closed orbit and the secondary closed orbit associated with the fixed point within the stable islands. A key criterion is that the difference between $\gamma_\mathrm{t}$ for the nominal closed orbit and that for the fixed point, $\gamma_\mathrm{t,nom}-\gamma_\mathrm{t,fp}$, must be sufficiently large for the specific situation under study. There are two scenarios to consider: one where $\gamma_\mathrm{t,fp} < \gamma_\mathrm{t,nom}$ and the other where the reverse is true.

\begin{figure}[htb]
\centering
  \includegraphics[trim=35truemm 30truemm 50truemm 25truemm,width=0.58\linewidth,angle=0,clip=]{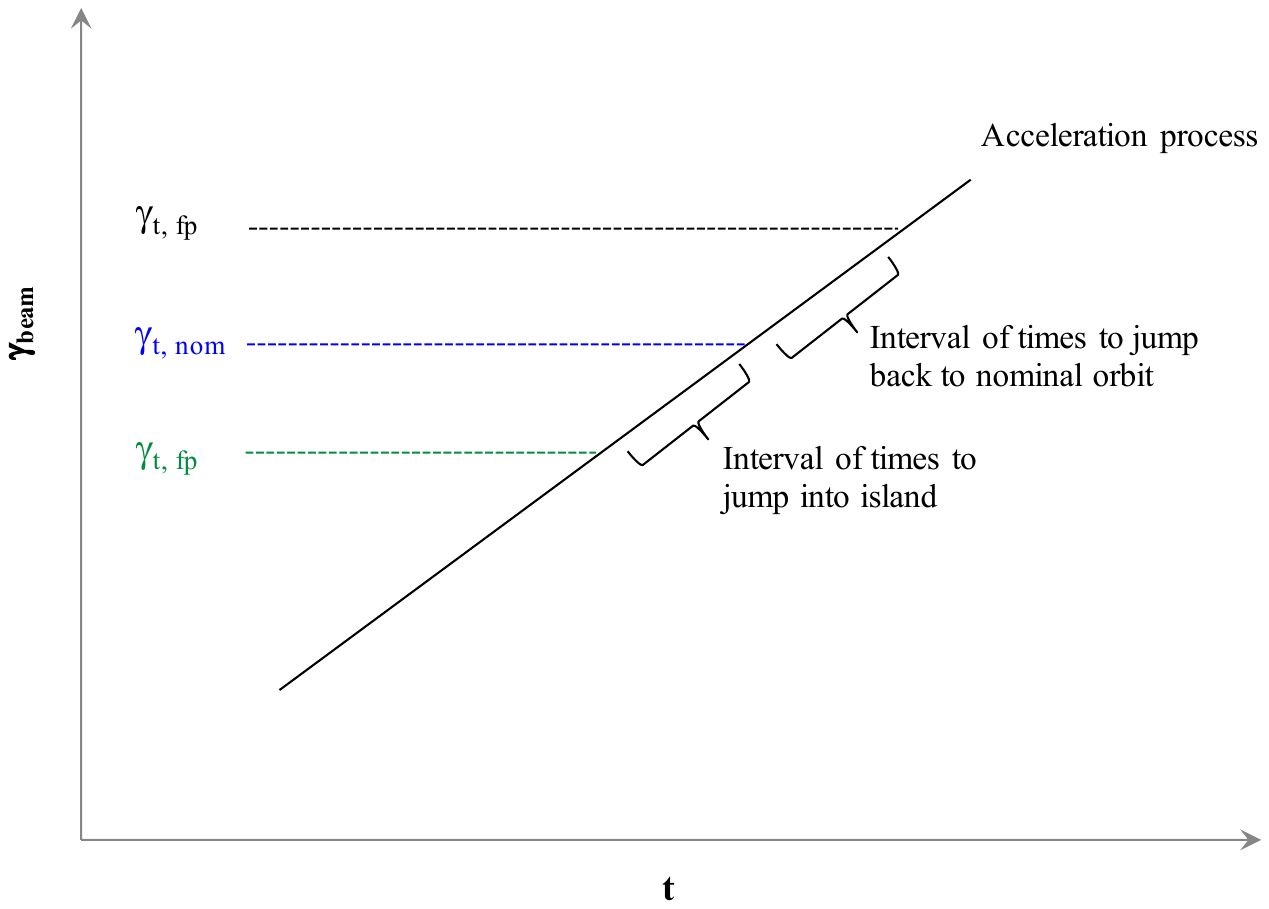}
  \caption{Outline of the proposed transition-crossing process. The time-dependent evolution of $\gamma_{\rm beam}$ is illustrated alongside the values of $\gamma_{\rm t, fp}$ (in green) and $\gamma_{\rm t, nom}$ (in blue), indicating the time span for transitioning into the island and subsequently returning to the standard closed orbit. Additionally, an alternative approach is presented for the scenario where $\gamma_{\rm t, fp}$ (in black) is larger than $\gamma_{\rm t, nom}$. The decision between these strategies should be guided by which setup is more straightforward to implement. (Adapted from Ref.~\cite{GiovannozziEPJPlusTransition}).}
\label{fig:sketch} 
\end{figure}

Stable islands can be formed and accessible in the horizontal phase space, but will initially remain unoccupied as the beam circulates along the nominal closed orbit. According to the first scenario described above, during acceleration, as $\gamma_\mathrm{beam}$ approaches $\gamma_\mathrm{t, nom}$ and falls within the range $\gamma_\mathrm{t,fp} < \gamma_\mathrm{beam} < \gamma_\mathrm{t,nom}$, a kicker can deflect the beam to a stable island. This beam, now on the secondary closed orbit, will be above the transition, ensuring that it is in a safe state and avoiding the transition associated with the standard closed orbit. Once $\gamma_\mathrm{beam}$ exceeds $\gamma_\mathrm{t,nom}$, the beam can be deflected back to the standard closed orbit with another dipole kick. The second scenario employs a similar technique, with the distinction that the beam's return to the standard closed orbit must be carefully timed to prevent the beam from undergoing the transition while within the stable islands. Figure~\ref{fig:sketch} outlines the proposed method for managing transition crossing.

The specific order of the resonance used does not affect the fundamental concept of the suggested transition-crossing technique. However, it is important to note that the area of the islands, which must be sufficiently large to accommodate the beam, decreases as the resonance order increases~\cite{Bazzani:262179}. Therefore, choosing a fourth-order resonance seems most advantageous.

A key benefit of the proposed method compared to the traditional gamma-jump approach is that it eliminates the requirement for pulsing magnets, aside from the kicker, which could be the same device that is used to inject or extract the beam. The proposed technique is based on non-linear beam dynamics and thus does not disturb the linear motion near the phase-space origin, leaving the beam dynamics around the standard closed orbit completely undisturbed.
\section{Conclusions} \label{sec:conc}
In this article, we reviewed the most significant applications of advanced non-linear dynamics concepts that have been developed under several different contexts: to describe the behaviour of particle beams in modern storage rings or colliders; to manipulate the transverse distributions of hadronic beams by crossing non-linear resonances in one or two dimensions; to generate stable secondary orbits that allow different optics in the same machine; to create beams that are longer than the length of the accelerator machine using stable resonances. 

Numerous applications have been examined theoretically with the aid of numerical simulations, using magnetic lattice models of accelerator rings with varying degrees of complexity. Some applications have involved practical experiments, and, in the case of the MTE beam for the CERN PS, the technology is even routinely employed for operation.

While these findings pave the way for numerous future applications, there remain certain facets that warrant thorough investigation to fully harness the potential offered by non-linear dynamics.

Since diffusive models have proven to be effective in explaining the evolution of a beam distribution function, identifying a method to compute the diffusion coefficient through numerical simulations has become crucial. This would facilitate the use of the Fokker-Planck formalism in tracking simulations of beam distributions over time scales compatible with those of the physical application. Furthermore, the entire formalism should be extended to describe the diffusive process in two degrees of freedom. 

The manipulation of the transverse beam distribution was initially envisioned under the assumption that charged particles do not interact. A series of experiments conducted at the CERN PS indicated that split beams are impacted by indirect space-charge effects~\cite{PhysRevSTAB.16.051001}, and these effects were validated through numerical simulations~\cite{PhysRevAccelBeams.20.121001}. The unique nature of distributions altered via non-linear effects adds complexity to numerical simulations, necessitating the development of new tools. This opens an extensive field of study into collective effects for multi-Gaussian distributions, encompassing both theoretical and numerical research. In addition to exploring space-charge effects, it would be highly valuable to investigate the behaviour of colliding split beams with the aim of developing new applications to enhance the performance of future colliders.

The implementation of magnetic components with oscillating fields appears to be a promising area for future developments. Potential applications could range from innovative particle-splitting methods to the cleaning of beam halos, a matter of significant interest for high-energy accelerators (see, e.g. Refs.~\cite{PhysRevLett.107.084802,stancari:napac13-tuoca1,redaelli:ipac15-webb1,PhysRevAccelBeams.23.031001,Redaelli_2021,PhysRevAccelBeams.24.021001} and references therein).

Finally, the domain of non-linear dynamics applied to a lepton circular accelerator is still in its infancy and remains to be fully explored, both theoretically, numerically and experimentally. 

Overall, non-linear beam dynamics has emerged as a highly productive area of accelerator physics, offering numerous opportunities for advancements in accelerator performance. It is anticipated that this will be an interesting field of study for researchers in the years ahead, given that there is now a willingness to move beyond the well-known and quiet realm of linear beam dynamics.
%
%
\section*{Acknowledgements}
The results of the intense research efforts presented in this article were made possible by the substantial input of many colleagues. I am sincerely grateful for past and ongoing fruitful collaborations with A.~Bazzani, F.~Capoani, A.~Franchi, S.~Gilardoni, C.~Hernalsteens, E.~H.~Maclean, C.~E.~Montanari, G.~Turchetti, F.~F~Van der Veken, and D.~E.~Veres.

I am grateful to F.~Zimmermann for his support and encouragement in the preparation of this article.

Lastly, I would like to express my deep gratitude to R. Cappi for involving me at the beginning of the discussions to devise a new type of lossless extraction to replace CT extraction. Without his intervention, I would not have had the opportunity to work and contribute to the exciting field of non-linear beam dynamics.
\bibliography{mybibliography}
\end{document}